\newif\ifreferee
\refereefalse
\documentclass{aa}

\usepackage{graphicx}
\usepackage{txfonts}
\usepackage{times}
\usepackage{epsfig}
\usepackage{graphicx}
\usepackage{amsmath}
\usepackage{amssymb}
\usepackage{tabulary}
\usepackage{multirow}

\begin{document}

\title{Deriving star cluster parameters with convolutional neural networks.}
\subtitle{II. Extinction and cluster/background classification}
\author{J. Bialopetravi\v{c}ius\inst{1}, D. Narbutis\inst{1, 2}}
\institute{
Vilnius University Observatory, Saul\.{e}tekio av. 3, LT-10257 Vilnius, Lithuania
\and
Center for Physical Sciences and Technology, Saul\.{e}tekio av. 3, LT-10257 Vilnius, Lithuania \\
\email{jonas.bialopetravicius\{@ff.vu.lt, @gmail.com\}}}

\abstract
{Convolutional neural networks (CNNs) have been established as the go-to method for fast object detection and classification on natural images. This opens the door for astrophysical parameter inference on the exponentially increasing amount of sky survey data. Until now, star cluster analysis was based on integral or resolved stellar photometry, which limits the amount of information that can be extracted from individual pixels of cluster images.}
{We aim to create a CNN capable of inferring star cluster evolutionary, structural, and environmental parameters from multi-band images, as well to demonstrate its capabilities in discriminating genuine clusters from galactic stellar backgrounds.}
{A CNN based on the deep residual network (ResNet) architecture was created and trained to infer cluster ages, masses, sizes, and extinctions, with respect to the degeneracies between them. Mock clusters placed on M83 Hubble Space Telescope (HST) images utilizing three photometric passbands (F336W, F438W, and F814W) were used. The CNN is also capable of predicting the likelihood of a cluster's presence in an image, as well as quantifying its visibility (signal-to-noise).}
{The CNN was tested on mock images of artificial clusters and has demonstrated reliable inference results for clusters of ages $\lesssim$100 Myr, extinctions $A_V$ between 0 and 3 mag, masses between $3\times10^3$ and $3\times10^5$ ${\rm M_\odot}$, and sizes between 0.04 and 0.4 arcsec at the distance of the M83 galaxy. Real M83 galaxy cluster parameter inference tests were performed with objects taken from previous studies and have demonstrated consistent results.}
{}

\keywords{methods: data analysis -- methods: statistical -- techniques: image processing -- galaxies: individual: M83 -- galaxies: star clusters: general}
\titlerunning{Deriving star cluster parameters with CNNs}
\authorrunning{Bialopetravi\v{c}ius and Narbutis}
\maketitle

\section{Introduction}

\begin{figure*}
    \centering
    \begin{tabular}{@{}c@{}}
        \includegraphics[width=0.48\textwidth]{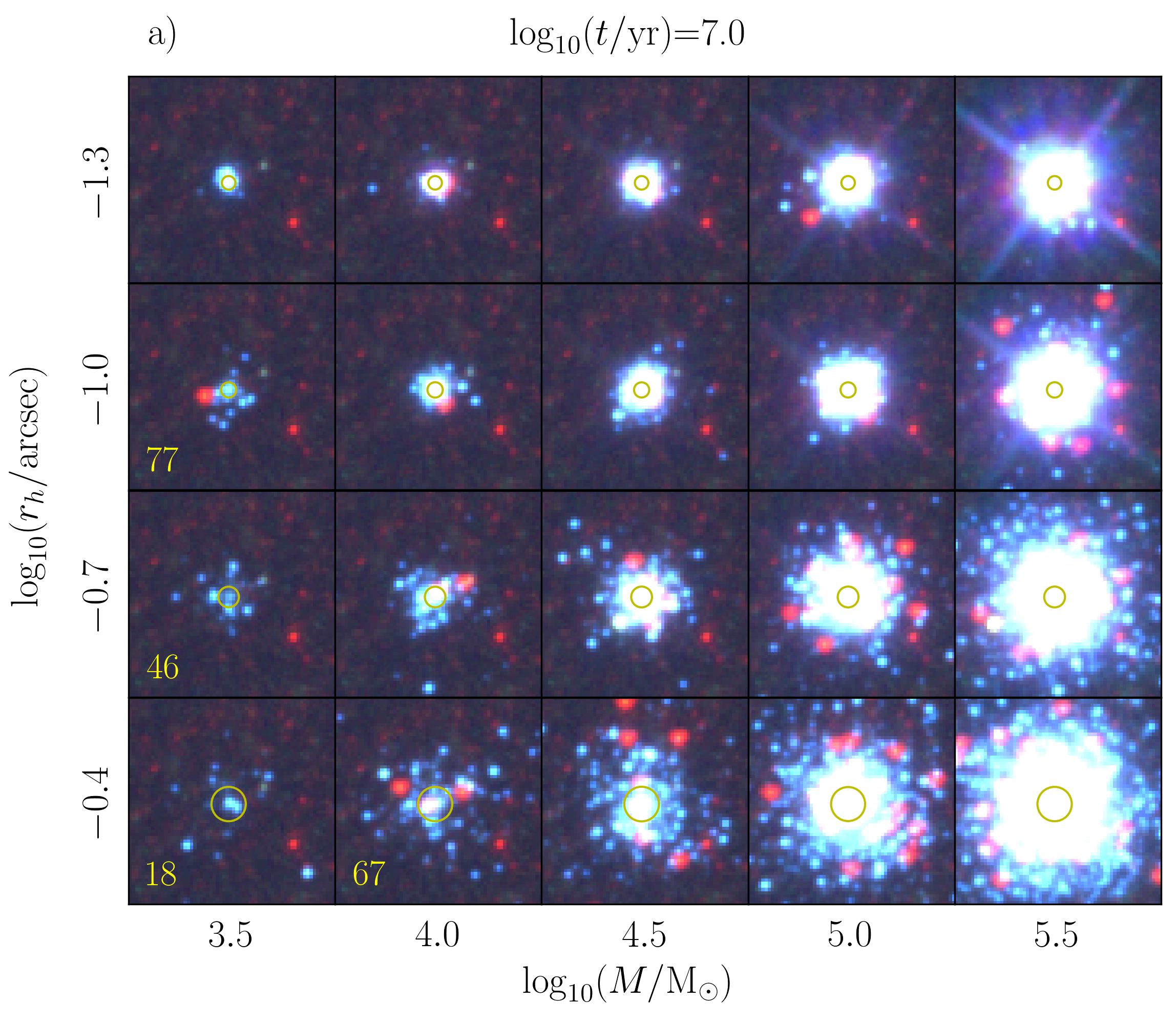}
        \\[0.5 pt]
    \end{tabular}
    \begin{tabular}{@{}c@{}}
        \includegraphics[width=0.48\textwidth]{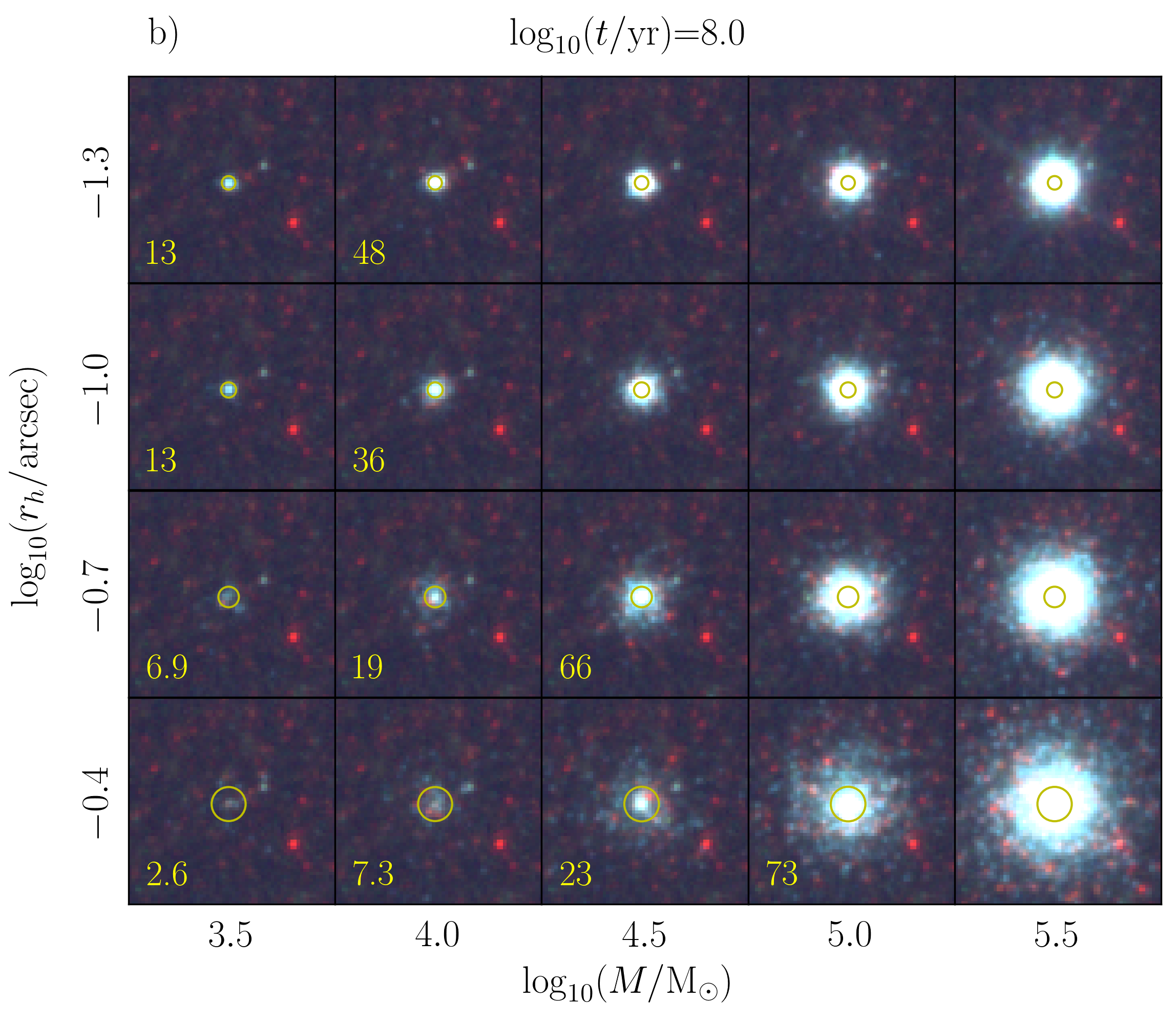} \\[0.5 pt]
    \end{tabular}
    \begin{tabular}{@{}c@{}}
        \includegraphics[width=0.48\textwidth]{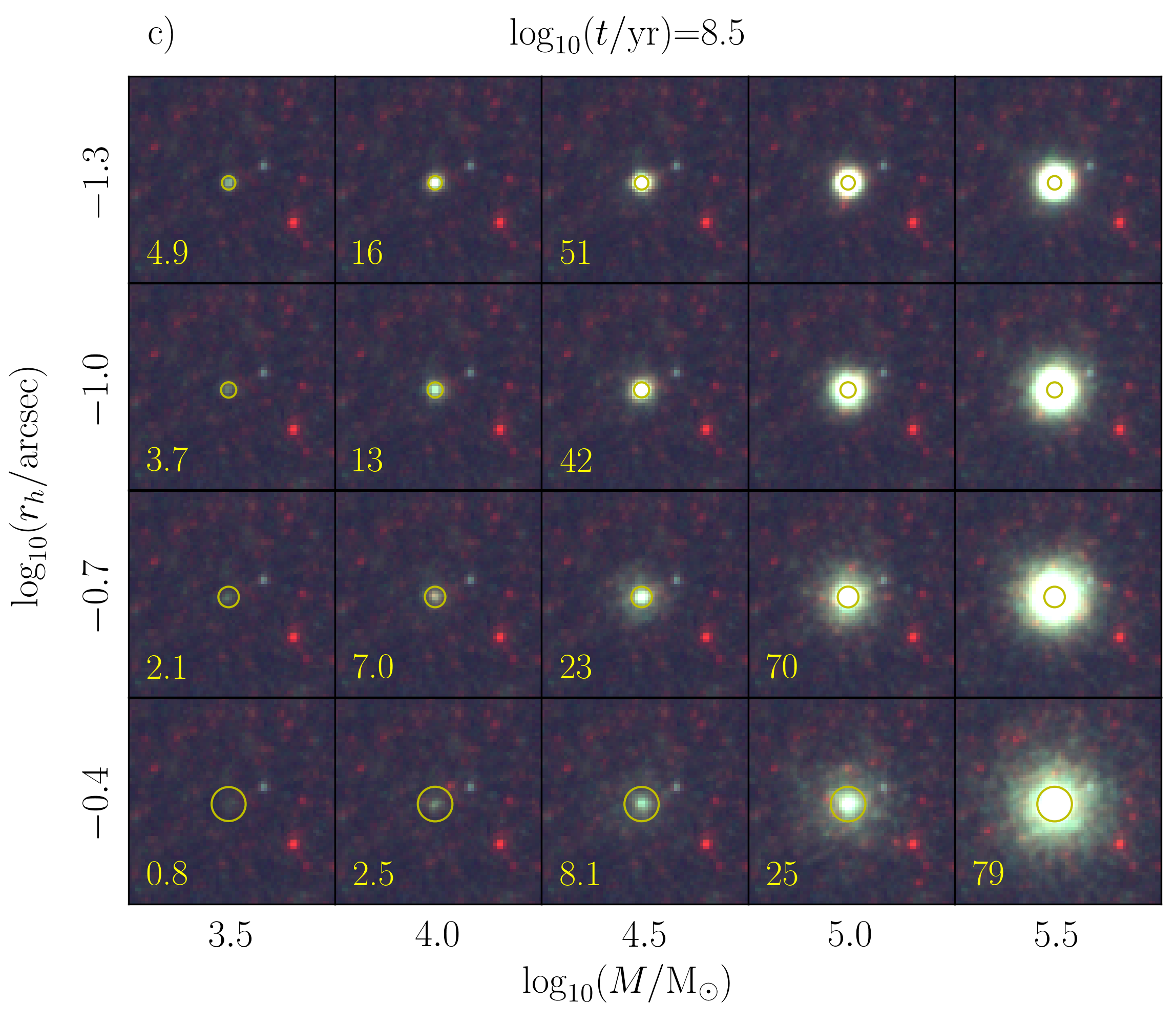} \\[0.5 pt]
    \end{tabular}
    \begin{tabular}{@{}c@{}}
        \includegraphics[width=0.48\textwidth]{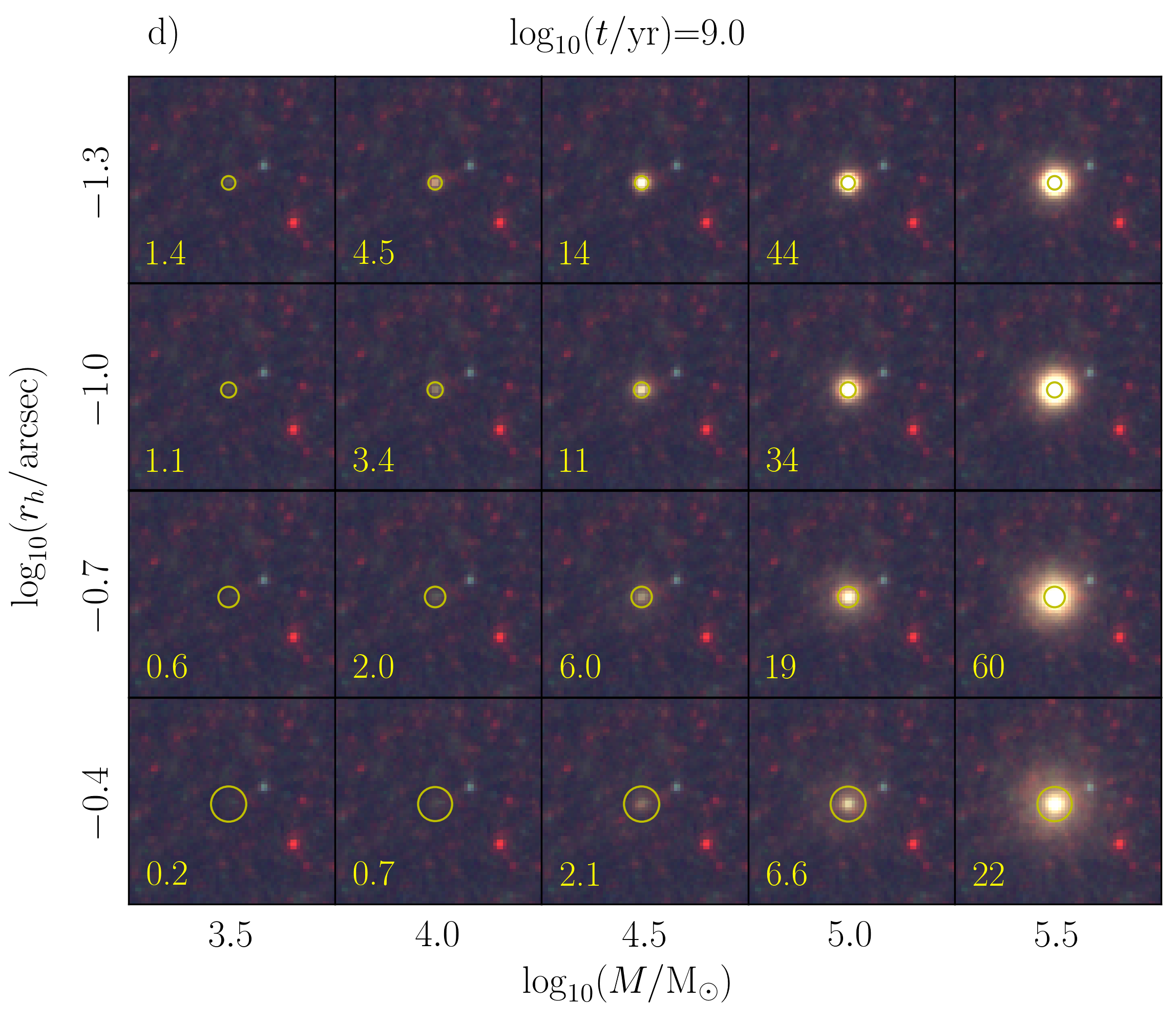} \\[0.5 pt]
    \end{tabular}
    \caption{Examples of generated mock clusters placed on a real background image, which is the same for all panels. The ages of the displayed clusters are shown on top of each panel. The mass and size, $r_h$, values are varied as shown on the axes; extinction $A_V=0$ mag. The intensity scale of the images was normalized with the arcsinh function within identical pixel value limits for each image. The yellow circles represent $r_{h,\,{\rm obs}}$ values ($r_h$ convolved with the point spread function). The $visibility$ (signal-to-noise proxy) value is displayed on the bottom-left of each image for fainter objects with $visibility<100$. Image sizes are $64\times64$ pixels ($2.6\times2.6$ arcsec) or $\sim$$60\times60$ pc at the distance of M83. In the images the color red corresponds to passband F814W, green to F438W, and blue to F336W.}
    \label{fig:artificial_cluster_samples_noextinction}
\end{figure*}

\begin{figure*}
    \centering
    \begin{tabular}{@{}c@{}}
        \includegraphics[width=0.48\textwidth]{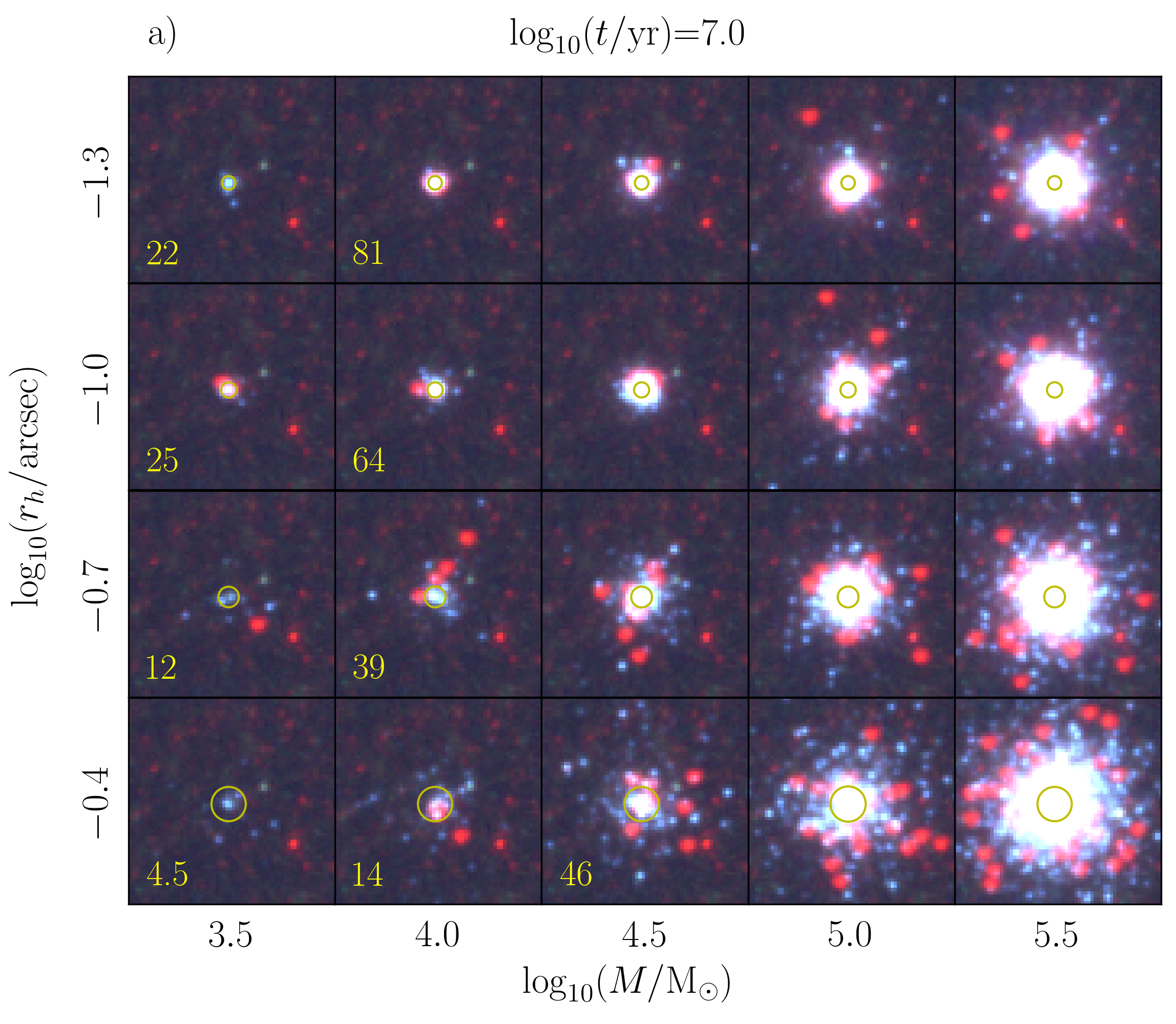} \\[0.5 pt]
    \end{tabular}
    \begin{tabular}{@{}c@{}}
        \includegraphics[width=0.48\textwidth]{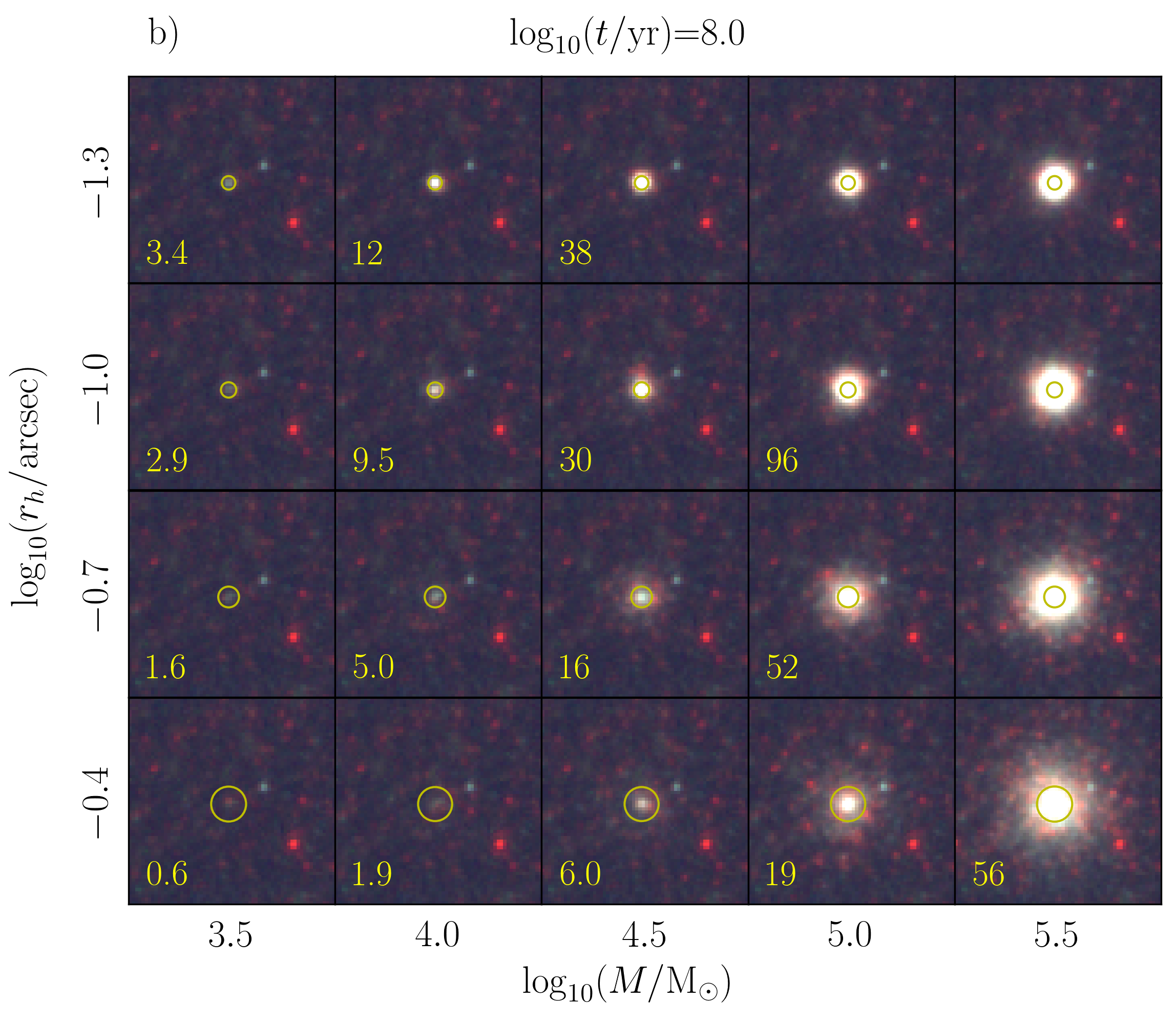} \\[0.5 pt]
    \end{tabular}
    \begin{tabular}{@{}c@{}}
        \includegraphics[width=0.48\textwidth]{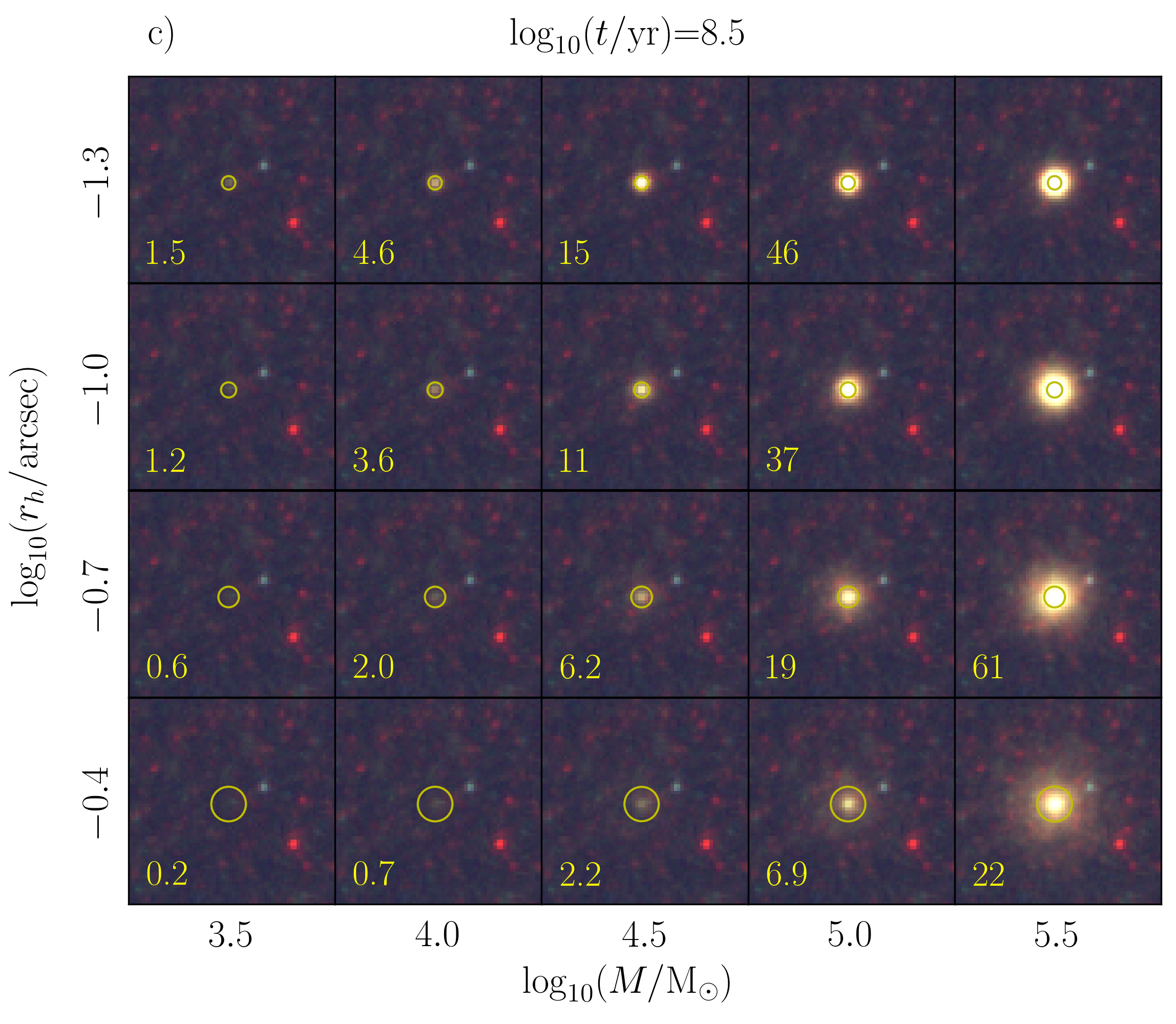} \\[0.5 pt]
    \end{tabular}
    \begin{tabular}{@{}c@{}}
        \includegraphics[width=0.48\textwidth]{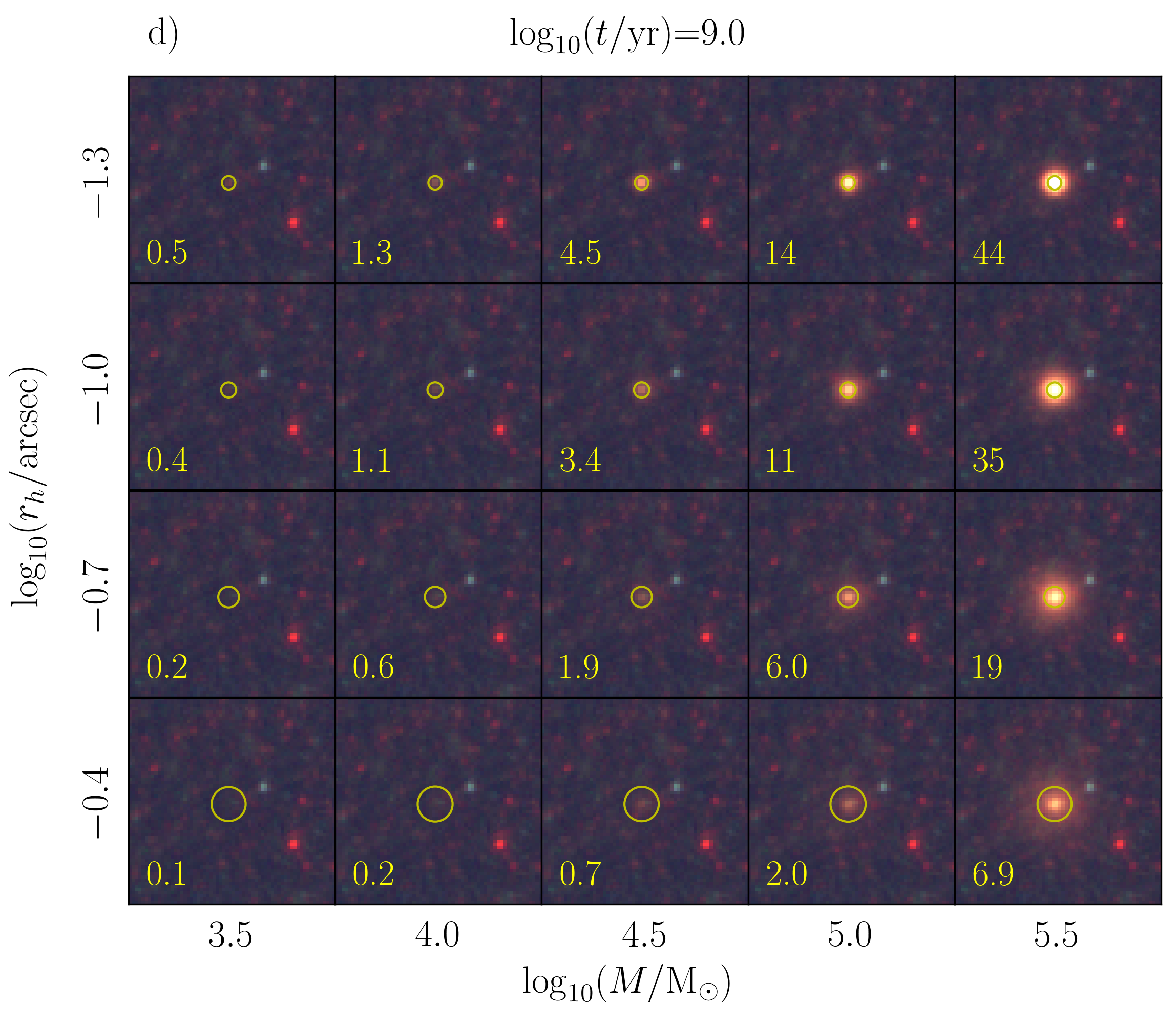} \\[0.5 pt]
    \end{tabular}
    \caption{Same as Fig. \ref{fig:artificial_cluster_samples_noextinction}, but with $A_V=1$ mag.}
    \label{fig:artificial_cluster_samples_extinction}
\end{figure*}

\begin{figure*}
    \centering
        \includegraphics[width=0.95\textwidth]{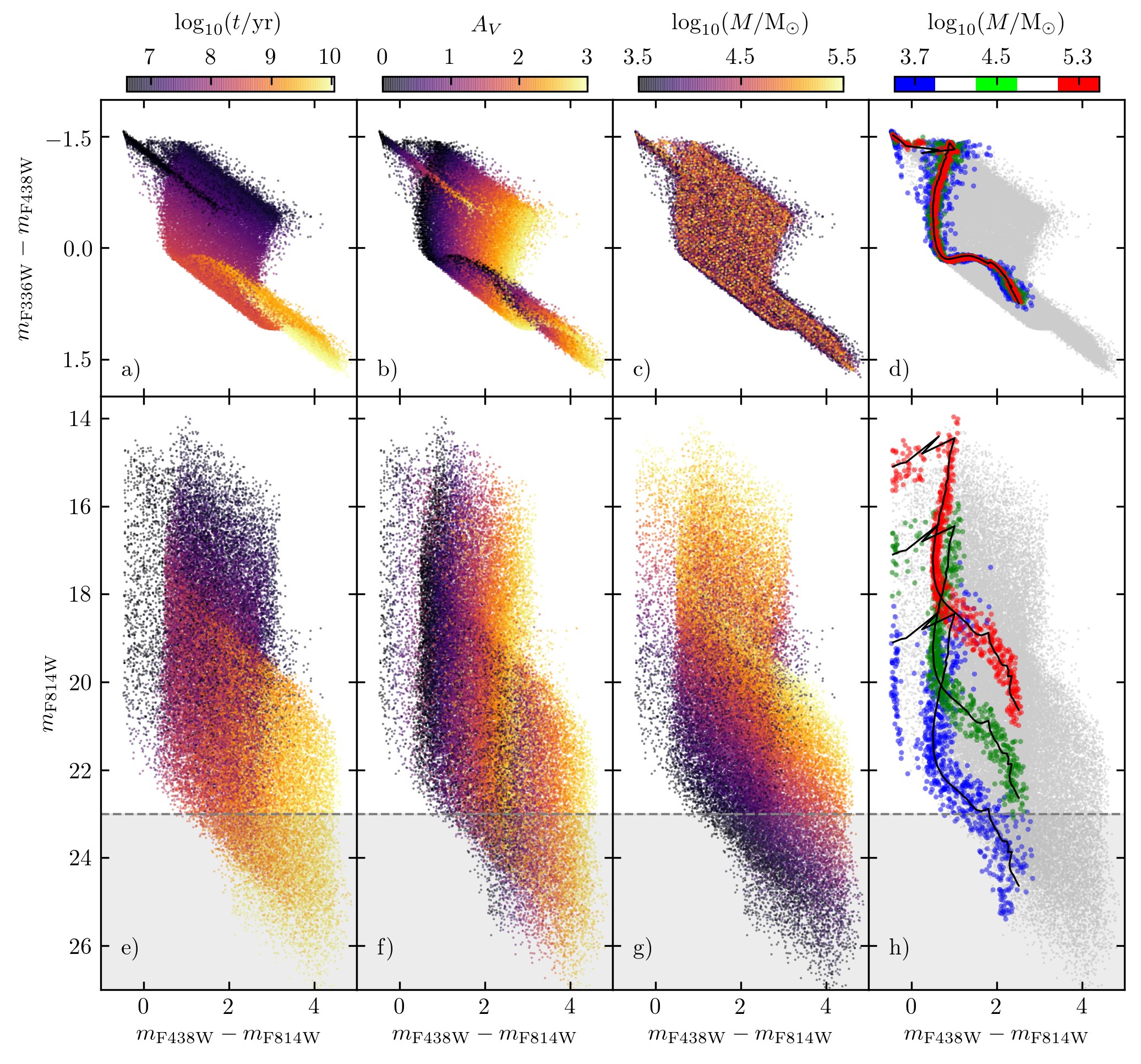}
    \caption{Integrated color-color and color-magnitude diagrams of 50,000 generated mock clusters of the training bank, as well as $\sim$7,000 faint clusters. The color coding represents different ages (a, e), extinctions (b, f), and masses (c, g). Their values are as noted on the color bars on top. The last column (d, h) shows distributions of $A_V<0.2$ mag star clusters filtered by three mass ranges as specified on the color bar on top. The simple stellar population tracks centered on the specified masses are shown as black curves. The shaded area below the dashed line represents the F814W magnitude limit used to filter out faint clusters.}
    \label{fig:mock_cmd}
\end{figure*}

\begin{figure}
    \centering
    \ifreferee
        \includegraphics[width=0.5\columnwidth]{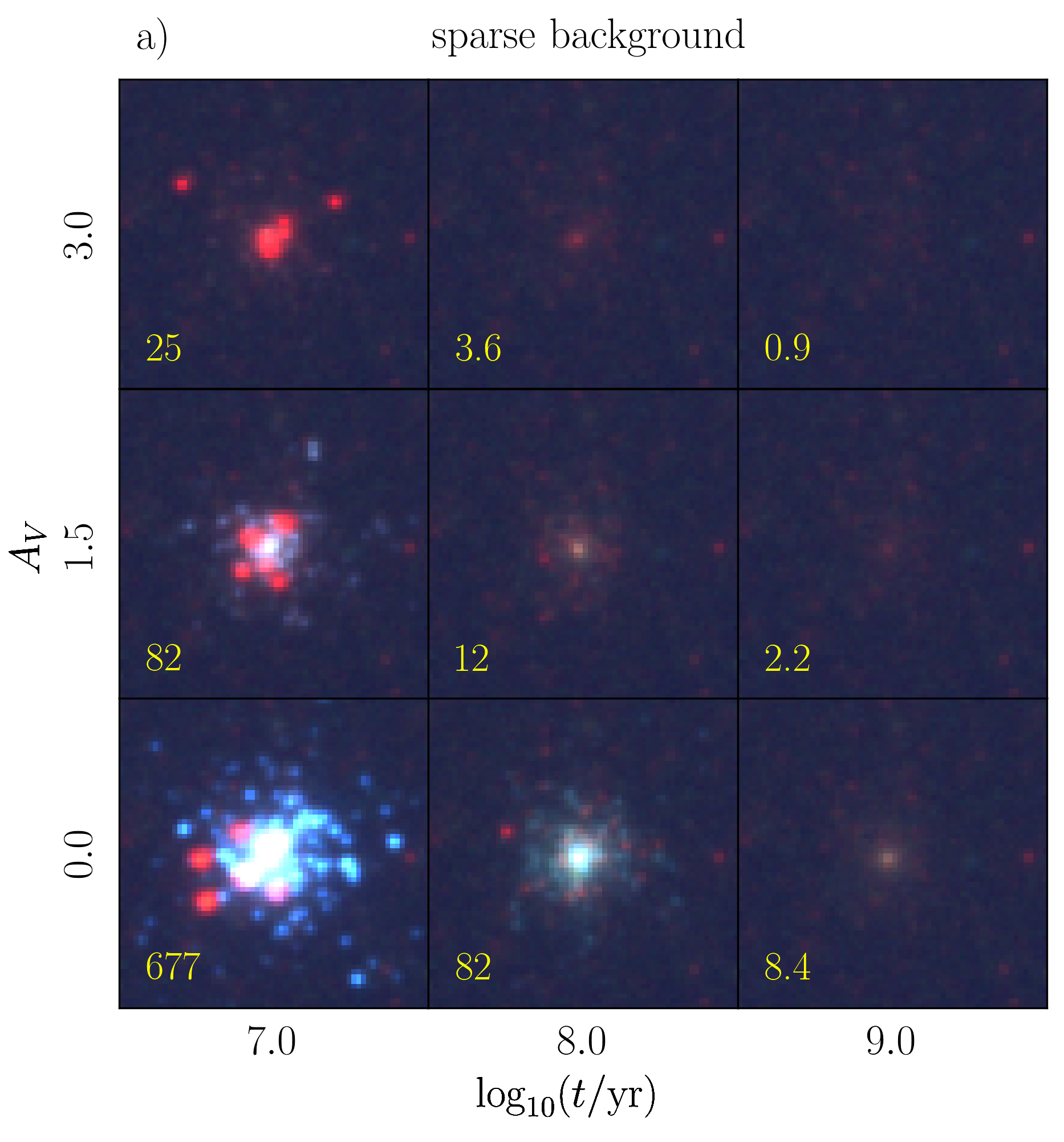} \\[0.5 pt]
        \includegraphics[width=0.5\columnwidth]{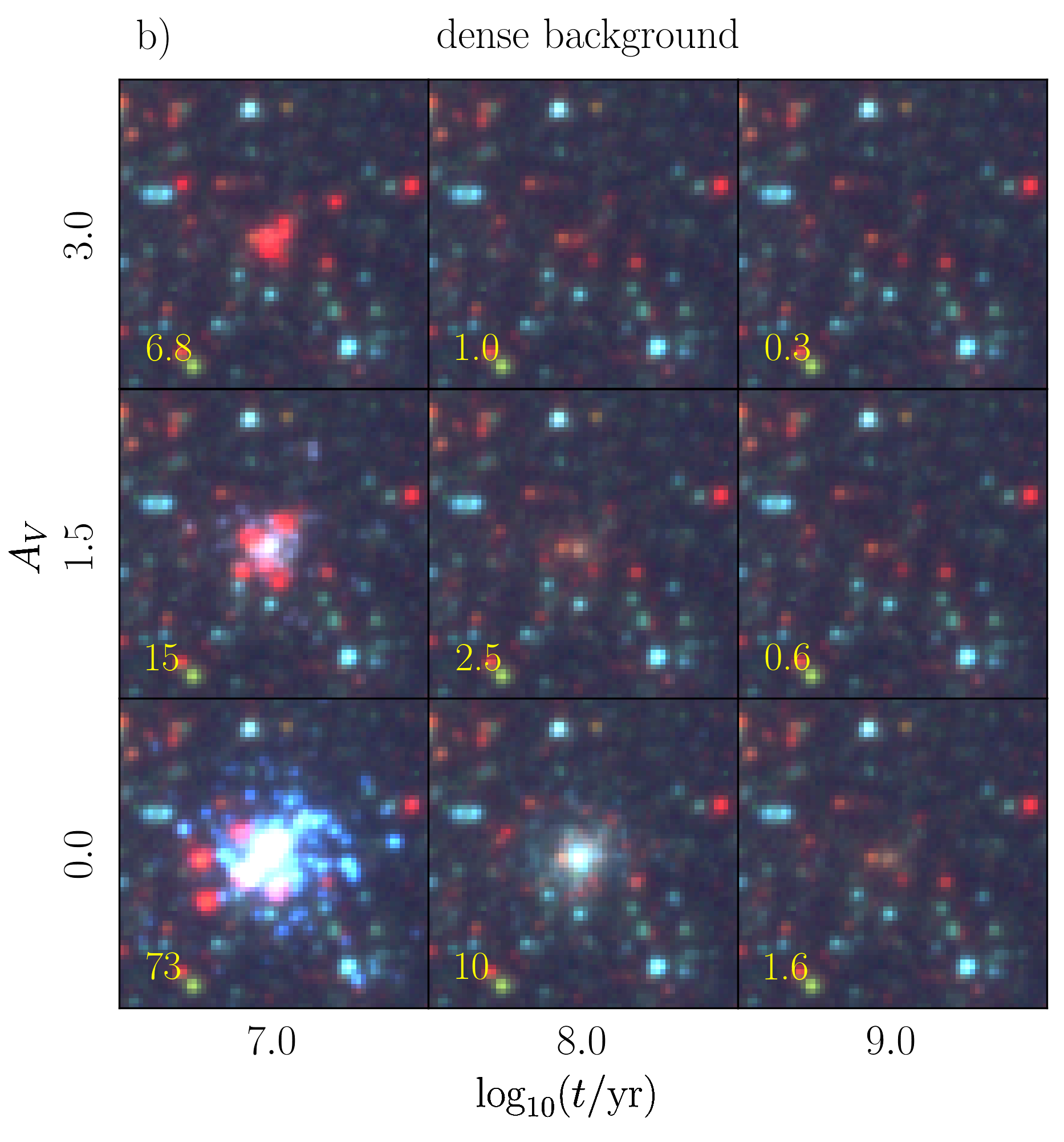} \\[0.5 pt]
    \else
        \includegraphics[width=0.9\columnwidth]{age_av_sample_sparse.png} \\[0.5 pt]
        \includegraphics[width=0.9\columnwidth]{age_av_sample_dense.png} \\[0.5 pt]
    \fi
    \caption{Examples of generated mock clusters with varying ages and extinctions on real background images. The masses of the clusters are $\log_{10}(M/{\rm M_\odot})=4.5$, their sizes are $\log_{10}(r_h/\rm{arcsec})=-0.7$. The images are normalized as in Fig. \ref{fig:artificial_cluster_samples_noextinction}. Top panel shows clusters superimposed on a sparse background, while the bottom panel shows the same clusters superimposed on a denser background. The $visibility$ value is displayed on the bottom-left of each image. The sizes of the images are $64\times64$ pix.}
    \label{fig:artificial_cluster_samples_ageav}
\end{figure}

Observational astronomy is a favorable field for computer vision applications and currently also experiences the accelerating uptake of convolutional neural networks (CNNs). These methods have drastically improved various object recognition tasks from natural images, such as object classification and detection \citep{ILSVRC15}. Just in the past year there have been numerous applications of CNNs in astrophysics, including galaxy shape estimation \citep{2019arXiv190208161R}, supernovae detection \citep{Reyes2018EnhancedRI}, and radio source morphology classification \citep{2019MNRAS.482.1211W}. Such progress strongly motivates the adaptation of CNNs in star cluster analysis. Furthermore, CNNs perform inference by processing all pixels of an image, which is beneficial for the parameter derivation task as demonstrated by \cite{2011ApJ...729...78W}, who used pixel-to-pixel variations to infer cluster ages.

In Paper I \citep{PaperI} we have implemented a CNN-based algorithm to simultaneously derive age, mass, and size of clusters in the low signal-to-noise regime. The algorithm was applied to M31 clusters, cataloged by The Panchromatic Hubble Andromeda Treasury (PHAT) survey. We have found that even when including information from all pixels and using accurate flux calibrations, interstellar extinction still plays a major role in influencing the results of parameter inference.

Numerous previous studies have explored physical parameter inference by taking into account the extinction problem, but were focused on the cases of resolved stellar or integrated cluster photometry. Among them are works by \cite{2008BaltA..17..337B}, who used analytically integrated stellar luminosities, \cite{2010A&A...521A..22F} and \cite{2013A&A...550A..20D,2014A&A...569A...4D}, who used stochastically sampled stellar luminosities according to the stellar initial mass function (IMF), and SLUG, developed by \cite{2015MNRAS.452.1447K}, which is one of the most mature codes in stochastic cluster population simulation and inference.

In this work we extend the CNN architecture proposed in Paper I to allow the inference of a cluster's interstellar extinction directly from images. With an eye towards automated star cluster detection, we also explore indicators of cluster presence in images. The outputs of the network were modified to infer multiple cluster parameters jointly, which allows the degeneracies between them to be expressed in the outputs of the network, instead of relying on single-point estimates. This is especially useful when visualizing and dealing with age-extinction degeneracies.

We used the M83 galaxy HST survey \citep{2014ApJ...788...55B}, which covers the entire disk of this face-on galaxy in a number of passbands. This allows us to investigate the effects of extinction in a variety of dense and sparse environments. Previous studies of the M83 star cluster population were based on aperture photometry, such as \cite{2015MNRAS.452..525R} covering the whole galactic disk, \cite{2011MNRAS.417L...6B} who studied a smaller part of the galaxy in detail, and \cite{Harris_2001} covering its central region.

We trained the CNN on realistic mock observations and tested on mock clusters, as well as validated on the aforementioned real cluster catalogs.

We also experimented with the re-normalization of image fluxes for each passband separately when training the network, suggesting that precise photometric calibrations may not be necessary to derive star cluster parameters. This was done in the vein of \cite{dieleman}, where JPEG color images were used to classify galaxies, achieving reliable results. This brings the approach of analysis of astronomical images closer to the methods used on natural images, which rarely have accurate flux calibrations.

The paper is organized as follows. Section \ref{sec:data_m83} provides details about the M83 survey data, the mock cluster bank construction, the added new parameters, and training data preparation. Section \ref{sec:cnn} describes the proposed CNN and its training methodology. Section \ref{sec:results_m83} presents the results of testing the method on mock as well as validating on real M83 clusters previously studied using integral photometry. Section \ref{sec:discussion} discusses the CNN parameter inference results in an astrophysical context.

\section{Data} \label{sec:data_m83}

\subsection{M83 mosaics}
The M83 mosaic project data observed by the HST Wide Field Camera 3 (WFC3) \citep{2014ApJ...788...55B,2010ApJ...710..964D} was obtained from the Mikulski Archive for Space Telescopes\footnote{https://archive.stsci.edu/prepds/m83mos/}. We use stacked, defect-free mosaic images of 7 WFC3 fields, which are calibrated photometrically (pixel values are in counts per second) and astrometrically (with available world coordinate system information). The details of image processing are provided by \cite{2014ApJ...788...55B}.

The mosaics cover the whole extent of the galaxy, from the dense center to its sparse outskirts where stellar background contamination is low. For the analysis we selected wide passband images that cover the whole galaxy without gaps: F336W, F438W, and F814W. All three mosaics are of the same size and in a tangential projection with a common scale (0.04 arcsec/pixel).

In Paper I the M31 images were masked for saturated stars and extended objects in order to prevent unreliable CNN training. The distance to the M31 galaxy is 785 kpc \citep{2005MNRAS.356..979M}, however the distance to M83 is 4.5 Mpc \citep{2003ApJ...590..256T}, therefore only a few saturated stars are visible. Because the area covered by extended objects in comparison to genuine stellar backgrounds is negligible, we decided to skip the masking step altogether and use all of the available mosaic area when selecting backgrounds for artificial clusters.

\subsection{Mock cluster generation}
\label{sec:artificial_clusters}

Mock clusters were generated with different ages, masses, sizes, and affected by various levels of extinction. A fixed metallicity of $Z=0.03$ \citep{2019ApJ...872..116H} and standard extinction law with $R_V = A_V/E(B-V) = 3.1$ were assumed. To generate a cluster, its parameters were sampled independently of each other either from continuous (for mass and $r_h$) or discrete (for age and $A_V$) ranges. For a cluster to be included into the bank it also had to be brighter than a defined magnitude limit as discussed below to create suitable data for network training. We note, that we do not perform grid sampling of cluster parameters, where all possible permutations of their discrete values are combined, which would be computationally expensive, and, as further results show, not necessary for the network to learn cluster features based on a limited number of examples.

The age of each cluster was drawn with a uniform probability from the logarithmic range of $\log(t/{\rm yr})=[6.6, 10.1]$ with a step of 0.05 dex, which corresponds to 71 discrete ages in the isochrone bank. Mass for each cluster was drawn with a uniform probability from the logarithmic range of $\log(M/{\rm M_\odot})=[3.5, 5.5]$ as a floating point number. The age and mass ranges were chosen in order to cover the majority of M83 clusters studied by \cite{2011MNRAS.417L...6B}. Extinction was drawn with a uniform probability from the range of $A_V=[0.0, 3.0]$ mag with a step of 0.1 mag, which corresponds to 31 discrete extinctions in the isochrone bank. We define $r_h$ as the radius of a circle on the sky enclosing half of the stars of a cluster. The spatial distributions of stars were drawn from the Elson-Fall-Freeman (EFF) \citep{1987ApJ...323...54E} profile:
\begin{equation}
\mu(r)=\mu_0(1+r^2/a^2)^{-\gamma/2}.
\end{equation}
The parameters $a$ and $\gamma$ were drawn with a uniform probability from logarithmic ranges of $[0.04, 1.2]$ and $[2.05, 8.0]$ respectively as floating point numbers, such that $r_h$ is within the limits of $[0.04, 0.4]$ arcsec. These values at the assumed distance of M83 \cite[4.5 Mpc]{2003ApJ...590..256T} roughly correspond to real cluster sizes ($r_h$) in M83 \citep{2011MNRAS.417L...6B}.

The stars of the clusters were generated as follows. Given the initial mass, $M$, of a cluster, star masses were sampled according to the \cite{2001MNRAS.322..231K} IMF from Padova PARSEC isochrones\footnote{http://stev.oapd.inaf.it/cgi-bin/cmd} \citep[release 1.2S]{2012MNRAS.427..127B}, obtaining the absolute star magnitudes for passbands F336W, F438W, and F814W. Then, the absolute magnitudes were transformed to apparent magnitudes at the distance of M83 \cite[4.5 Mpc]{2003ApJ...590..256T} and converted to the WFC3 camera counts per second for the three passbands using calibrations provided by \cite{2012wfci.book.....D}. Finally, the spatial 2D positions of stars were generated by sampling their distances from the cluster's center according to the EFF profile (with given $a$ and $\gamma$ values) and then distributing them symmetrically around the center.

The GalSim package \citep{2015A&C....10..121R} was used to draw the individual stars of the clusters using TinyTim-generated\footnote{http://tinytim.stsci.edu/cgi-bin/tinytimweb.cgi} point spread functions (PSFs) \citep{2011SPIE.8127E..0JK} for each of the three passbands. Every star in the cluster was drawn separately for each passband using the appropriate PSF scaled by the star's flux in counts per second. For a single cluster this produces three images, which can then be visualized as either RGB pictures or given to a CNN as 3D (width $\times$ height $\times$ passband) arrays. Artificial clusters were then placed on backgrounds cut from the M83 mosaics. See Figs. \ref{fig:artificial_cluster_samples_noextinction} and \ref{fig:artificial_cluster_samples_extinction} for examples of the generated mock clusters.

To explore the photometric properties of the cluster bank, we show integrated color-color and color-magnitude diagrams in Fig. \ref{fig:mock_cmd}. The magnitudes depicted were obtained solely from integrating the total flux of mock clusters and therefore are an idealized case, which does not take into account the variations of background and spatial positions of stars. The only source of stochastic effects in such a case is IMF sampling. Panels are dedicated to illustrate the influence of age, extinction, and mass present in the bank. The effects of these parameters are in different directions in the color-color and color-magnitude space. The oldest clusters are red (panel a) and low-luminosity (panel e) objects. Clusters with high extinction are reddened (panel b), and the lowest mass clusters are faintest (panel g).

The last column (Fig. \ref{fig:mock_cmd}, panels d and h) shows distributions of star clusters filtered by mass (as specified on the color bar on top) and by extinction $A_V<0.2$ mag. The simple stellar population (SSP) tracks centered on the specified masses are shown as black curves. In both, color-color and color-magnitude space, it can be seen that lower mass clusters are more widely distributed due to the stochastic IMF sampling. The effects of mass on cluster magnitude can be seen again in Fig. \ref{fig:mock_cmd}h as vertical shifts of the SSP tracks.

This means that a point in color-color and color-magnitude space can't uniquely map to a point in cluster parameter space. This is worsened by stochastic IMF sampling effects and results in degeneracies with which any parameter inference method has to deal with. In cases like this any additional sources of information, such as individual image pixel values, are welcome.

Faint objects with $m_{\rm F336W}>24$ mag, $m_{\rm F438W}>23.5$ mag, and $m_{\rm F814W}>23$ mag were not included in the final cluster bank due to their low signal, to mimic age/mass/extinction selection effects existing in magnitude limited real cluster samples. As adding these extremely faint clusters to real backgrounds would result in mock images that are below the detection limit, the CNN would be forced to learn a cluster's parameters on what effectively is just a plain background image. Therefore, magnitude cuts applied are necessary to provide the CNN with a balanced dataset. For the F814W band this is illustrated by the shaded gray area in Fig. \ref{fig:mock_cmd}. See the lower-left corner of panel d in Figs. \ref{fig:artificial_cluster_samples_noextinction} and \ref{fig:artificial_cluster_samples_extinction} for examples of such barely visible clusters.

\subsection{Mock cluster properties}

Samples of artificial clusters were generated with the described parameters and placed on real backgrounds of M83. In order to realistically model photon noise the following steps were applied. A cutout image of an M83 background from a random position in the mosaics is selected and its median value is determined. This median is then added to the image of an artificial cluster, multiplied by the exposure time to get photon counts, and then each pixel is sampled from a Poisson distribution, with its mean set to the value of the pixel. The median is then subtracted back from this image, the real background image is added and photon counts are transformed back to counts per second.

We also define a cluster $visibility$ parameter, constructing it to approximate signal-to-noise in such a way that higher values would be assigned to clusters that stand out relative to their stochastic stellar backgrounds. It is defined as follows:
\begin{equation}
visibility=\frac{f_c}{n\cdot\sigma_{b}},
\end{equation}
where $f_c$ is the integral flux of the cluster within its $r_{h,\,{\rm obs}}$, while $\sigma_{b}$ is the standard deviation of the background's pixel values in a 25 pix (1 arcsec) radius aperture, and $n$ is the number of pixels within $r_{h,\,{\rm obs}}$. Here $r_{h,\,{\rm obs}}$ is the cluster's $r_h$ value increased to account for PSF size, which has the largest effect on the most compact clusters. A mock cluster with $visibility = 1$ has mean flux per pixel approximately equal to the value of the standard deviation of a background it is placed on.

\begin{figure}
    \centering
    \ifreferee
        \includegraphics[width=0.4\columnwidth]{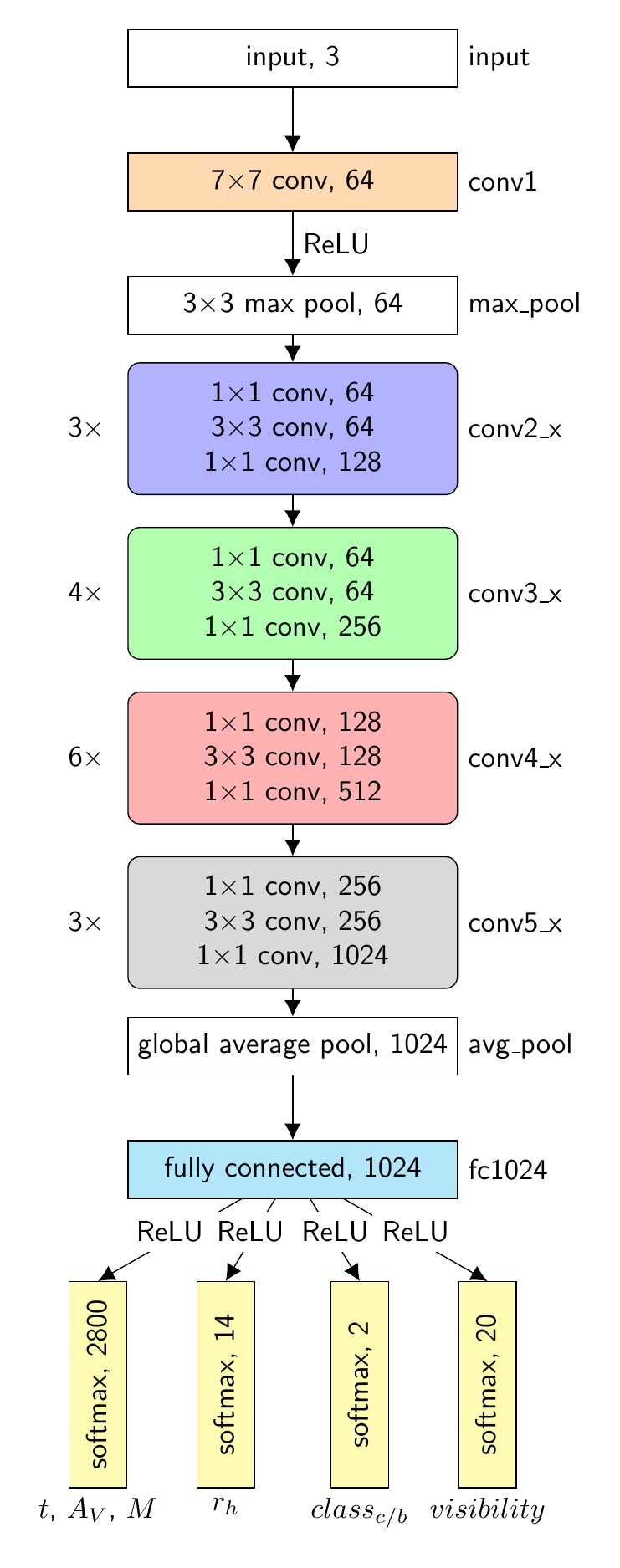}
    \else
        \includegraphics[width=0.75\columnwidth]{network_graph.pdf}
    \fi
    \caption{A block diagram of the CNN. The three-channel input image of a cluster passes through the network top to bottom, with the output result being age, extinction, mass, size, the cluster/background class $class_{c/b}$, and $visibility$. All blocks with sharp corners depict single layers, while blocks with rounded corners are groupings of layers, with the number on the left indicating how many times the group is repeated sequentially and the name on the right corresponding to the layer names in Fig. \ref{fig:network_architecture_table2}. The blocks with non-white backgrounds are parts of the network with optimizable parameters. The last number in each row is the number of output channels from that layer. ``ReLU'' indicates the locations in the network where rectified linear activations are applied between blocks.}
    \label{fig:network_architecture_graph2}
\end{figure}

\begin{figure}
    \centering
    \ifreferee
        \includegraphics[width=0.5\columnwidth,trim={6.8cm 17.7cm 6.8cm 1.5cm},clip]{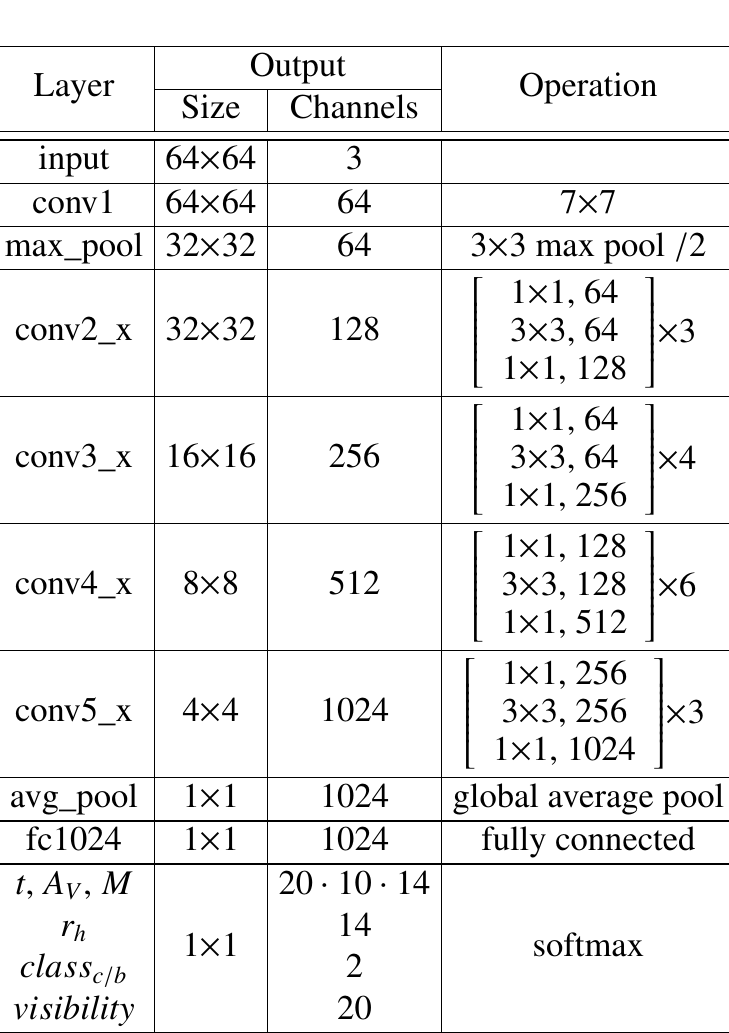}
    \else
        \includegraphics[width=0.8\columnwidth,trim={6.8cm 17.7cm 6.8cm 1.5cm},clip]{arch.pdf}
    \fi
    \caption{The designed 50-layer CNN, based on the ResNet architecture. The layers of the network are listed top to bottom, starting from the images of clusters and with the final layer producing cluster's parameters. The convolutional layers are actually groups of blocks, with the ``\_x'' in the name acting as a placeholder for the block number. The size of the outputs of each layer, both in spatial dimensions and in channel count, are listed on the second and third columns. The last column lists the operations that each layer performs. The layers or blocks with a stride of 2 are: max\_pool, conv3\_1, conv4\_1, and conv5\_1; as can be seen when input and output sizes differ 2 times. The last layer has 4 groups of softmax layers branching out in parallel, with the first predicting age, extinction, and mass, the second -- cluster size, the third -- the cluster/background class, and the fourth -- $visibility$.}
    \label{fig:network_architecture_table2}
\end{figure}

See Figs. \ref{fig:artificial_cluster_samples_noextinction} and \ref{fig:artificial_cluster_samples_extinction} for a variety of $visibility$ values of clusters, displayed as yellow text in the corner of each image, with $A_V$ up to 1 mag. See Fig. \ref{fig:artificial_cluster_samples_ageav} for samples of clusters with the full range of extinction ($A_V$ up to 3 mag) used in this study to illustrate the effect of background crowding on $visibility$. It can be seen that the values of $visibility$ correlate well with the ability to resolve clusters by eye -- the best tool for cluster detection up to date.

Note, that for real clusters it is not possible to infer properties of background covered by cluster's light, however by placing mock objects into backgrounds, we can compute $visibility$ parameter beforehand and train the network to infer it from the data of real observations.

\subsection{Training data preparation}
To minimize the influence of photometric image calibration accuracy, the counts per second of each passband of a cluster's image were individually normalized to the mean of 0 and standard deviation of 1. They were then rescaled with the arcsinh function. The resulting images were $64\times64$ pixels in size, which correspond to $2.6\times2.6$ arcsec, or $60\times60$ pc at the distance of M83 \cite[4.5 Mpc]{2003ApJ...590..256T}. Examples of the generated clusters with different ages, masses, and sizes, and without extinction, covering most of the parameter space, are shown in Fig. \ref{fig:artificial_cluster_samples_noextinction}. A series of different examples (star position and mass sampling), but with extinction $A_V=1$ mag, are shown in Fig. \ref{fig:artificial_cluster_samples_extinction}. We generated 50,000 such images of mock clusters as a training sample for the CNN. The backgrounds have also been precomputed for efficiency resulting in 80,000 cutouts that were combined with the cluster images.

\section{Convolutional Neural Network} \label{sec:cnn}
\subsection{Architecture} \label{sec:convolutional_neural_networks}

Following the work in Paper I, the ResNet-50 \citep{2015arXiv151203385H} architecture was used as a basis for our CNN. In addition, a series of modifications were made to it in order to accommodate the different survey images, the higher number of predicted parameters, as well as the degeneracies between them. See Figs. \ref{fig:network_architecture_graph2} and \ref{fig:network_architecture_table2} for details on the structure of the modified CNN.

\begin{figure}
    \centering
    \ifreferee
        \includegraphics[width=0.6\columnwidth]{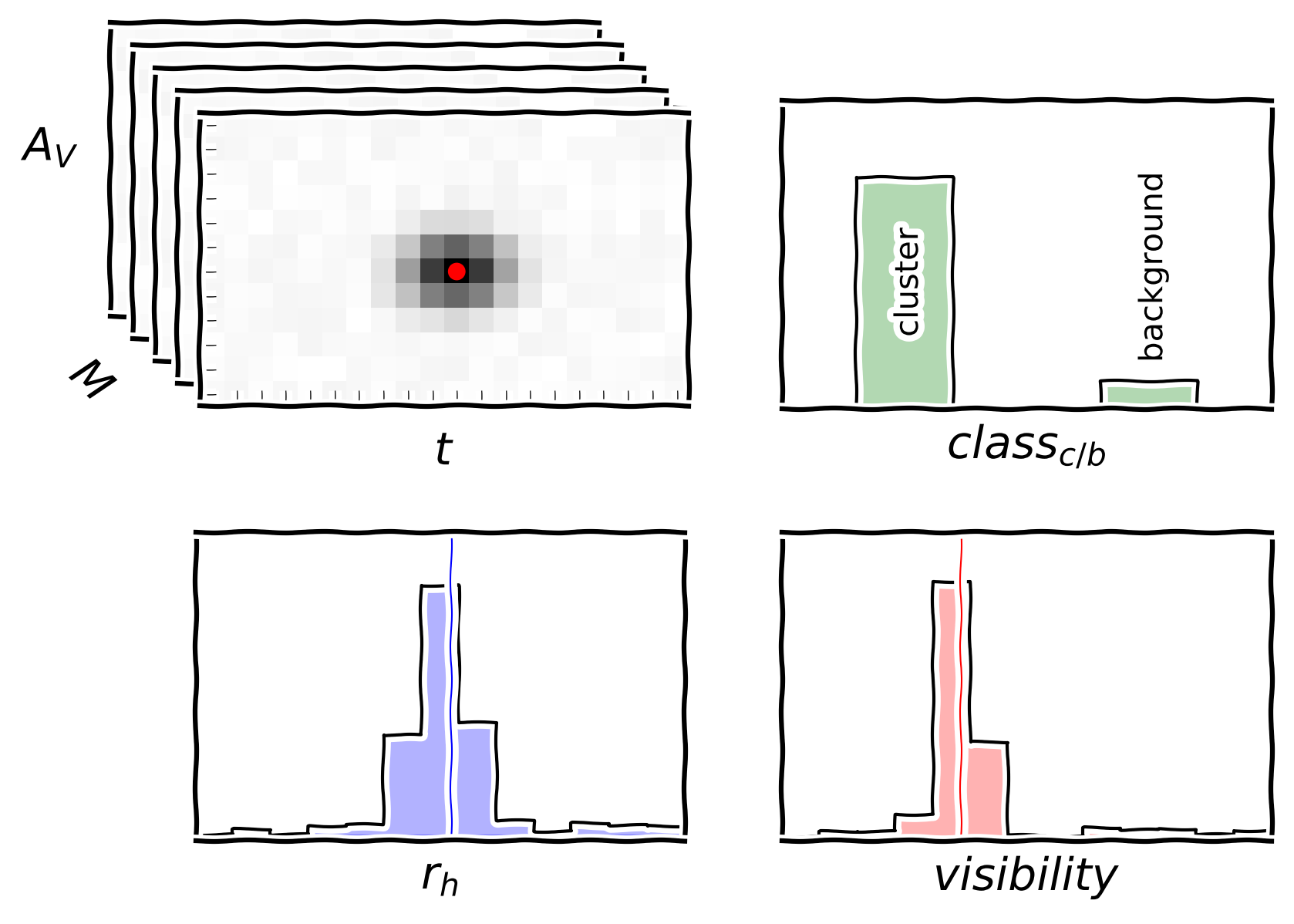}
    \else
        \includegraphics[width=0.95\columnwidth]{cnn_outputs.png}
    \fi
    \caption{An illustration of the activations in the output layers of the CNN. While training the CNN, target activation values are provided as a 3D Gaussian distribution for age, extinction, and mass (centered on true values as denoted by the red dot) and as 1D Gaussian distributions for $r_h$ (the blue line) and $visibility$ (the red line). The cluster/background class is represented as a value of 0 or 1. During inference the network produces similar outputs, examples of which are depicted in Figs. \ref{fig:stochastic_effects} and \ref{fig:example_inference_results}.}
    \label{fig:cnn_outputs}
\end{figure}

In Paper I we used a method by \citet{dieleman} to rotate the input image multiple times and pass it through the same convolutional layers; to simplify the network we omitted this step. The input image size was decreased to 64$\times$64 pixels to account for the smaller angular size of the clusters due to the more distant galaxy. Three input channels were used corresponding to the F336W, F438W, and F814W passbands.

In Paper I, the cluster's parameters were predicted via linear output layers by treating it as a regression problem. This meant that each parameter was predicted independently. However, due to age/extinction degeneracies and age/extinction/mass selection effects (shown in Fig. \ref{fig:mock_cmd}) this approach is no longer viable.

Therefore, we predict all of the parameters on a grid, with the positions on it corresponding to the parameter values. This essentially transforms the regression problem into classification, allowing the network to predict each parameter in multiple locations of the parameter space, properly representing some degenerate cases such as low-extinction and old-age being just as likely as high-extinction and young-age.

The network's output are 4 groups of layers branching out in parallel. The first group predicts age, extinction, and mass, the second -- cluster size, the third -- cluster/background class ($class_{c/b}$), and the fourth -- cluster $visibility$ (see bottom of Fig. \ref{fig:network_architecture_graph2}).

Fig. \ref{fig:cnn_outputs} depicts the four output layer activations. We grouped age, extinction, and mass into a single output layer to allow the degeneracies between these parameters to be expressed in the network architecture itself. This was done by predicting them as activations on a 3D grid, with 20 bins for age, 10 for extinction, and 14 for mass. When flattened, this results in a softmax layer with $K=2800$ neurons. For $class_{c/b}$ $K=2$ neurons were used to encode the likelihood of a cluster's presence in the image. For the remaining parameters single-dimensional grids were used, resulting in $K=14$ neurons for size and $K=20$ for $visibility$.

Each of the four groups of output parameters were represented as softmax activations:
\begin{equation} \label{eq:softmax}
\sigma(\vec{z})_i=\frac{e^{z_i}}{\sum_{j=1}^K e^{z_j}},
\end{equation}
where $\vec{z}$ is the activations of a whole layer, and $i$ specifies the index of a neuron (position on the parameter grid). The network was implemented with Keras\footnote{https://keras.io/} and TensorFlow\footnote{https://www.tensorflow.org/} packages.

\subsection{Training and inference}
\label{sec:training}

When training the network, we wish to infer both, $class_{c/b}$, which indicates the presence of a cluster, and the cluster's astrophysical parameters, at the same time. To that end learning the $class_{c/b}$ parameter was modeled as a simple binary classification task. The network is trained on batches of 512 images, half of which are images of backgrounds, and the other half are images of backgrounds combined with clusters as described in Section \ref{sec:artificial_clusters}. For the images with only background in them we set $class_{c/b}=0$, while for the samples with clusters we set $class_{c/b}=1$.

\begin{figure}
    \centering
    \ifreferee
        \includegraphics[width=0.5\columnwidth]{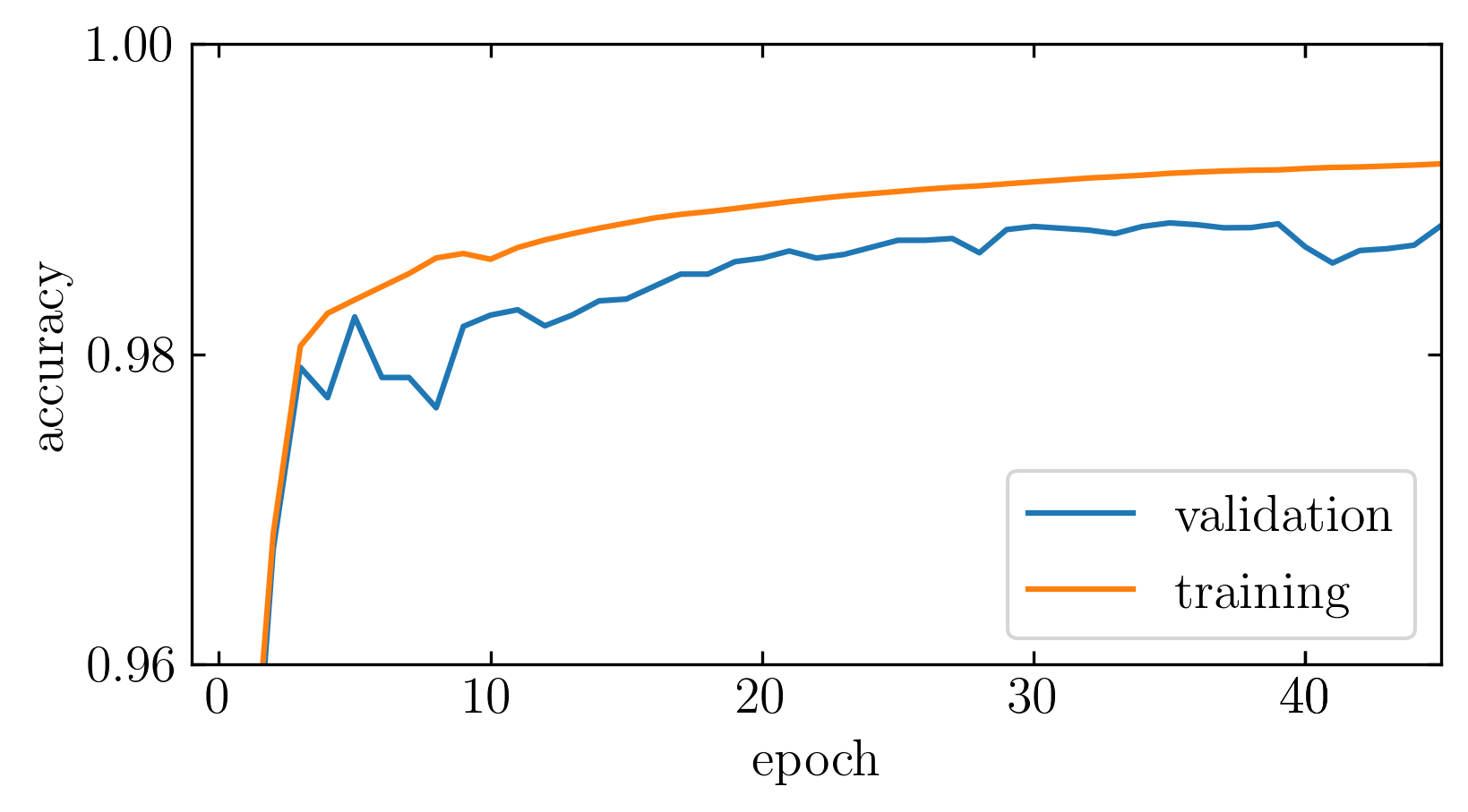}
    \else
        \includegraphics[width=0.8\columnwidth]{training_metrics.png}
    \fi
    \caption{Binary classification accuracy metric for $class_{c/b}$ on the training (orange) and validation (blue) sets during the course of CNN training.}
    \label{fig:loss}
\end{figure}

In addition, for background images we zero out the training loss gradients for all cluster parameters. In effect this causes gradient updates to only be derived from the $class_{c/b}$ and $visibility$ parameters, both of which are set to 0, indicating that the background contains no cluster. Training proceeds by sampling from M83 backgrounds ($\sim$25,000 images) and the cluster bank (50,000 mock clusters) separately, combining the cluster and background images on the fly, effectively giving us over $10^9$ unique training samples.

The usual way to encode real-valued parameters as bins is called one-hot encoding. The parameter space is divided into bins and the bin at the position of the parameter's value is set to 1. This array is then passed as a target vector, $\vec{y}$, for the network. One-hot encoding is ideal for categorical classification, where only one of the target bins is true at a time. However, for binned real-valued parameters this has the unfortunate side-effect of penalizing bins far away from the target just as much as bins nearby to it. The way we solve this is by inserting a Gaussian distribution centered on the true value of the parameter (see Fig. \ref{fig:cnn_outputs}). For the case of $r_h$ and $visibility$ this is a simple 1D Gaussian, with a standard deviation equal to 0.5 the width of a bin. For age, mass, and $A_V$ a 3D Gaussian was used, with a standard deviation equal to 0.25 the width of a bin.

\begin{figure*}
    \centering
    \includegraphics[width=0.435\textwidth]{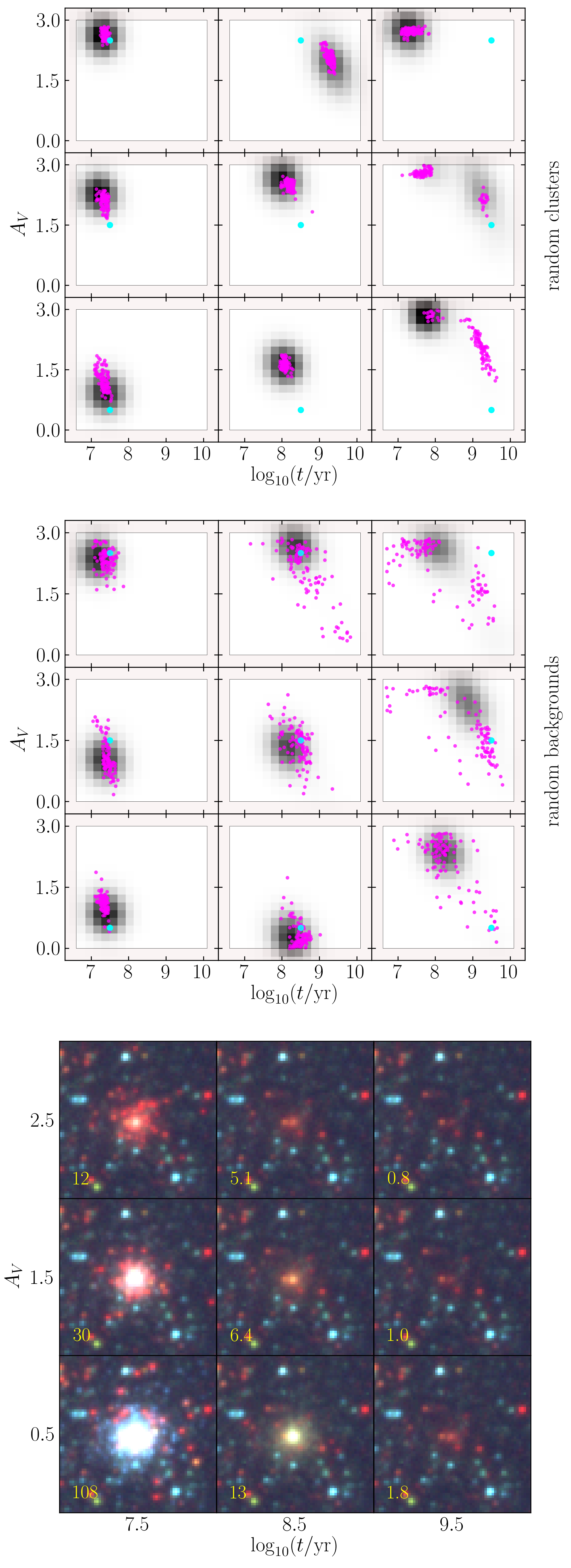}
    \hspace{0.6cm}
    \includegraphics[width=0.435\textwidth]{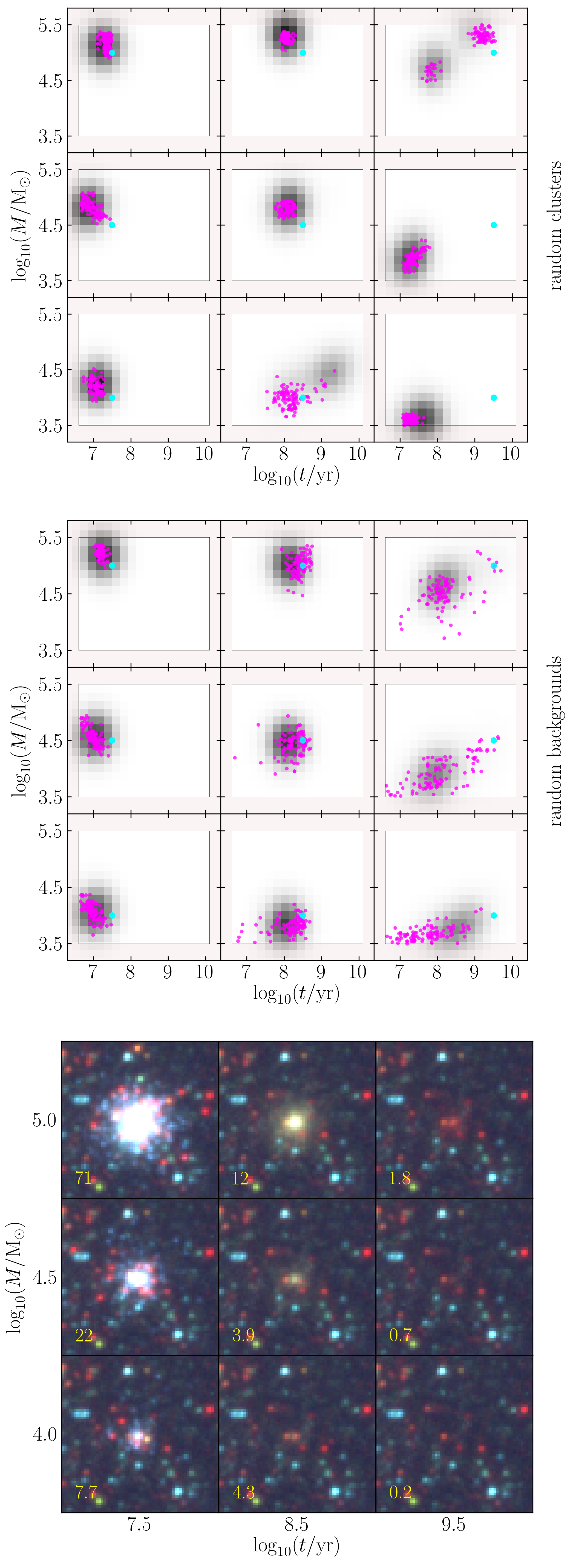}
    \caption{Influence of stochastic effects on mock cluster inference results. Left column shows clusters with $\log_{10}(M/{\rm M_\odot})=5.0$ and varied extinction and age. Right column shows clusters with $A_V=0.5$ mag and varied mass and age. Cluster sizes are fixed at $\log_{10}(r_h/\rm{arcsec})=-0.6$ for all cases. The top row shows the results of inference when stellar IMF sampling and spatial positions are varied while holding the cluster parameters constant. The middle row shows the results of inference when background images are varied while using the same cluster image. The cyan circles correspond to the true values of parameters. The grayscale colormaps are raw CNN outputs (activations over the parameter space) for one specific mock cluster case, while magenta circles show 100 single-point estimates obtained for different random cases. The bottom row shows visualizations of the clusters with fixed background in the same format as in Fig. \ref{fig:artificial_cluster_samples_ageav}.}
    \label{fig:stochastic_effects}
\end{figure*}

To obtain parameter estimates from this network we need a way to transform the network's output activations back into single-point estimates that can then be analyzed. The 1D and 3D histograms, depicted in Fig. \ref{fig:cnn_outputs}, need to be ``unfolded''. This was done by finding the bin with the highest value in the histogram, which represents the most likely set of parameters inferred by the network, and calculating a weighted average within a radius of 3 bin widths. In effect this produces an output that is a real-valued single-point estimate in-between the bins instead of a discrete-valued one. Examples of inference results on mock and real clusters with both, the raw activation outputs and the derived single-point estimates, are shown in Figs. \ref{fig:stochastic_effects} and \ref{fig:example_inference_results}.

For computing the training gradients for the network the categorical cross-entropy loss function was used:
\begin{equation}
\mathcal{L}=-\sum_{i=1}^K y_{i}\log(\sigma(\vec{z})_i),
\end{equation}
where $\sigma(\vec{z})_i$ is the neuron's activation as described in Eq. \ref{eq:softmax}, and $y_{i}$ is the target output for the given training cluster image.

The Adam optimizer \citep{2014arXiv1412.6980K} was used to calculate the gradients at each step of training. We experimented with various learning rates, starting from $0.1$ down to $0.0001$, with the learning rate decaying down to $0.0001$ at the final iteration of training for all experiments. The learning rate of $0.01$ gave the best performance on the mock validation set, so this was the value used for the final training of the network. The best CNN model was selected by picking the training iteration during which the CNN's loss was the lowest on the validation set. The training accuracy track of the $class_{c/b}$ parameter for the resulting model is shown in Fig. \ref{fig:loss}. This is the only parameter for which accuracy can be meaningfully calculated, because the other parameters are encoded as Gaussian distributions.

\subsection{Output activations and stochastic effects} \label{sec:stochastic_effects}

Three types of stochastic effects play a major role in the variation of CNN-inferred cluster parameters: 1) stellar mass sampling, 2) star spatial position randomization, and 3) background field.

We combine stellar mass and position sampling into one stochastic factor as a property of the cluster itself, while leaving the background choice as a property of its environment. We study both effects separately by: a) generating 100 different clusters with fixed parameters and placing them on the same background, and b) placing the same cluster on 100 different backgrounds.

Fig. \ref{fig:stochastic_effects} displays the influence of stochastic effects on the inference results of mock clusters. Left column shows clusters with $\log_{10}(M/{\rm M_\odot})=5.0$, extinctions $A_V$ = 0.5, 1.5, 2.5 mag, and $\log(t/{\rm yr})$ = 7.5, 8.5, 9.5. Right column shows clusters with $A_V=0.5$ mag, $\log_{10}(M/{\rm M_\odot})$ = 4.0, 4.5, 5.0, and the same ages. Cluster sizes are fixed at $\log_{10}(r_h/\rm{arcsec})=-0.6$ for all cases. The top row shows the results of inference when stellar IMF sampling and spatial positions are varied while holding the cluster parameters constant. The middle row shows the results of inference when background images are varied while using the same cluster image. The cyan circles correspond to the true values of parameters. The grayscale colormaps are raw CNN outputs for one specific case and magenta circles show 100 single-point estimates obtained for different random cases. The bottom row shows visualizations of the clusters with fixed background in the same format as Fig. \ref{fig:artificial_cluster_samples_ageav}. The $visibility$ parameter value is displayed on the bottom-left of each image. Note that the CNN predicts ages, masses, and extinctions as one 3D cube, while the outputs shown here are marginalized either over mass (left column) or extinction (right column).

In Fig. \ref{fig:stochastic_effects} it can be seen that the inference results for clusters with high visibility are all tightly packed for both types of stochastic effects (top and middle rows). This applies for both the spread of the CNN activation maps (grayscale) as well as the single-point estimates on different cluster images (magenta dots).

However, as clusters get fainter, and especially when they disappear into the background, the spread of activation maps (grayscale) as well as single-point estimates (magenta dots) gets wider. Background variability has a significantly larger influence on the spread of parameter estimates than stellar sampling effects.

It is worth noting that for old clusters CNN output activations are elongated, attempting to represent age/extinction degeneracies. For a small number of cases bimodal solutions are obtained. However, for cases where clusters are completely invisible both activations and single-point estimates can end up tightly packed. This highlights the importance of the $visibility$ parameter.

We note that less than 1\% of the mock clusters show the bimodal distribution of activations. About 20\% of the mock sample shows an extended unimodal distribution, while the rest of the results are symmetric and unimodal. Therefore, selecting the highest activation and obtaining single-point estimates from it is a viable approach, as that captures most of the information present in the CNN outputs.

For some clusters a systematic bias of the inference results can be observed, where the spread of activations would not sufficiently explain cluster misclassifications. This mainly occurs for the barely-visible clusters, where only cluster sampling is varied, implying that with a sufficiently difficult background and for a faint cluster, its parameter estimates may not be reliable and uncertainties may be underestimated.

However, Fig. \ref{fig:stochastic_effects} also illustrates the possibility to quantify the uncertainties of single inference results either from the extent of activation maps or by sampling random backgrounds, adding them to a cluster's image, and re-running inference. The former can produce tightly packed (underestimated) activation maps for some high-uncertainty samples, making them unreliable for low-visibility scenarios. The latter can also introduce additional effects, depending on the used background sampling method, as well as the tendency to overestimate the uncertainties on real clusters, since the background effects would get doubled.

In subsequent sections the single-point estimates are analyzed in respect to inferred parameter accuracy and the age/extinction degeneracy.

\section{Results} \label{sec:results_m83}
\subsection{Tests on mock clusters}

\begin{figure}
    \centering
    \ifreferee
        \includegraphics[width=0.65\columnwidth]{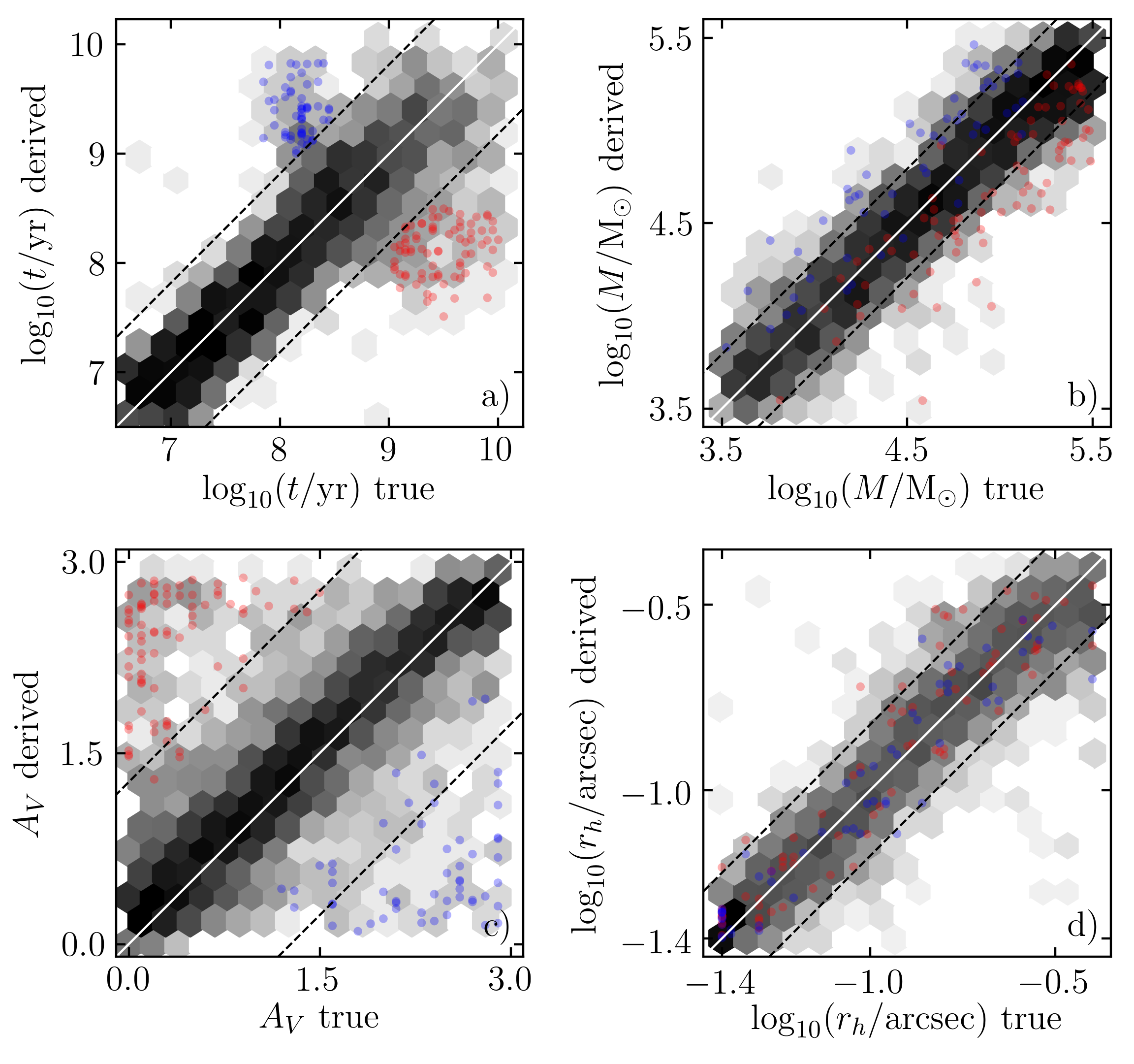}
    \else
        \includegraphics[width=1\columnwidth]{diagonal_mock_differences.png}
    \fi
    \caption{The true and derived parameter values of test mock clusters visualized as a hexagonal density map. The bins are scaled logarithmically. Panels show comparisons for a) age, b) mass, c) $A_V$, and d) $r_h$. Dashed lines highlight the area containing 95\% of the clusters. Red dots represent the clusters that were misclassified as younger than the real age values, while blue dots represent clusters misclassified as older.}
    \label{fig:mock_diagonal_results}
\end{figure}

To test the performance of the CNN, we built a separate bank of 5,000 artificial clusters. Their parameters were drawn from the same distributions as described in Section \ref{sec:artificial_clusters}. The backgrounds for these mock clusters were also sampled from the used M83 mosaic, making sure that they are not the same as the backgrounds used for training. The inferred parameter values were obtained as described in Section \ref{sec:training}.

Differences between CNN-derived single-point estimates of age, mass, extinction, and size vs. true parameters are shown in Fig. \ref{fig:mock_diagonal_results}. The spread of errors is visualized as a hexagonal density map with the count bins scaled logarithmically in order to highlight the spread of outliers. Dashed lines represent the error bounds containing 95\% of the inference results for each parameter. Note that because of magnitude cuts introduced in the mock cluster bank, discussed in Section \ref{sec:artificial_clusters}, the parameter distributions aren't uniform. For example, there are relatively less low-mass old-age clusters. In all of the panels, the clusters that are classified as much younger than the true given values are shown as red points, while the clusters classified as much older are highlighted as blue.

Fig. \ref{fig:mock_diagonal_results}a shows no significant difference between the true and derived age values for $\log_{10}(t/{\rm yr})<8$ and the distribution for all ages is symmetrical along the diagonal. The 95\% of all inference results deviate $<$0.9 dex from the true values, as shown by the dashed lines. Starting at $\log_{10}(t/{\rm yr})=8$ and above a large scatter in both directions -- towards older and younger ages can be seen.

\begin{figure}
    \centering
    \ifreferee
        \includegraphics[width=0.65\columnwidth]{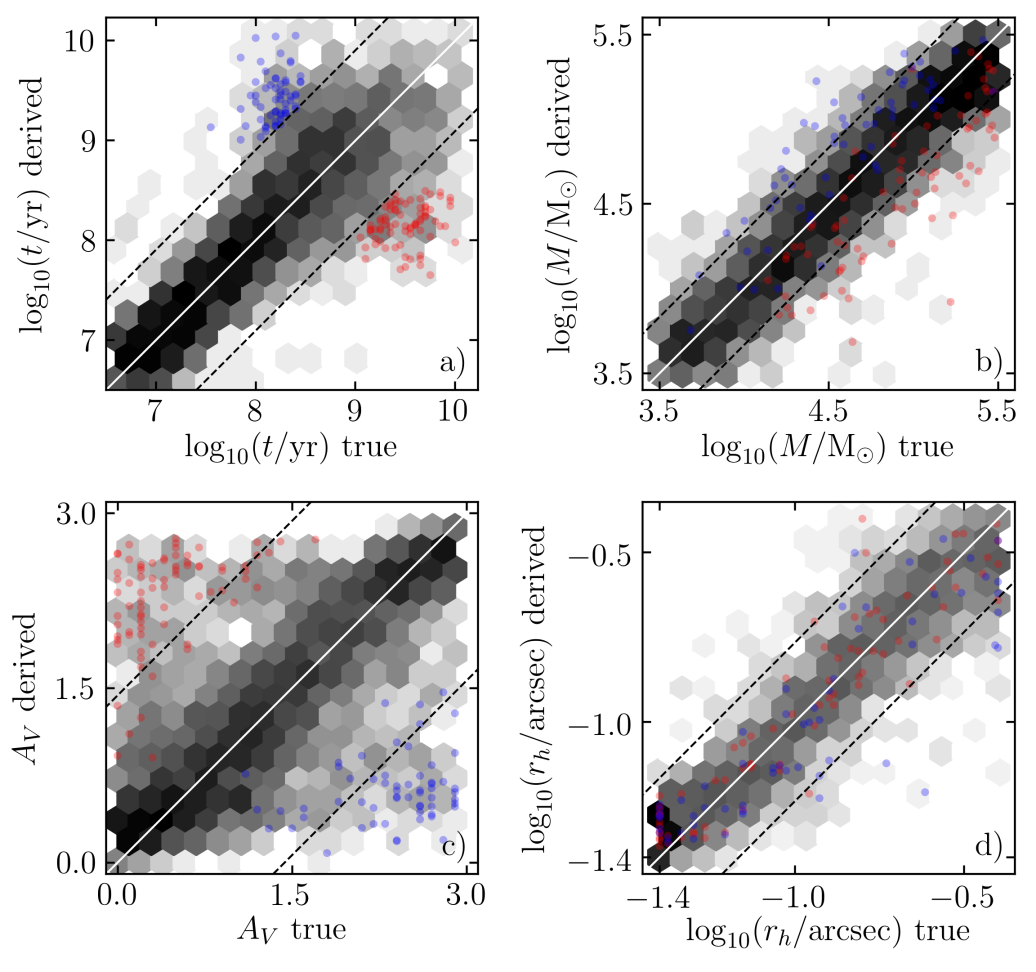}
    \else
        \includegraphics[width=1\columnwidth]{diagonal_mock_differences_varbg.png}
    \fi
    \caption{Same as Fig. \ref{fig:mock_diagonal_results}, but using a CNN which was trained on mock clusters with simulated uncertainty of photometric calibrations.}
    \label{fig:mock_diagonal_results_varbg}
\end{figure}

\begin{figure}
    \centering
    \ifreferee
        \includegraphics[width=0.65\columnwidth]{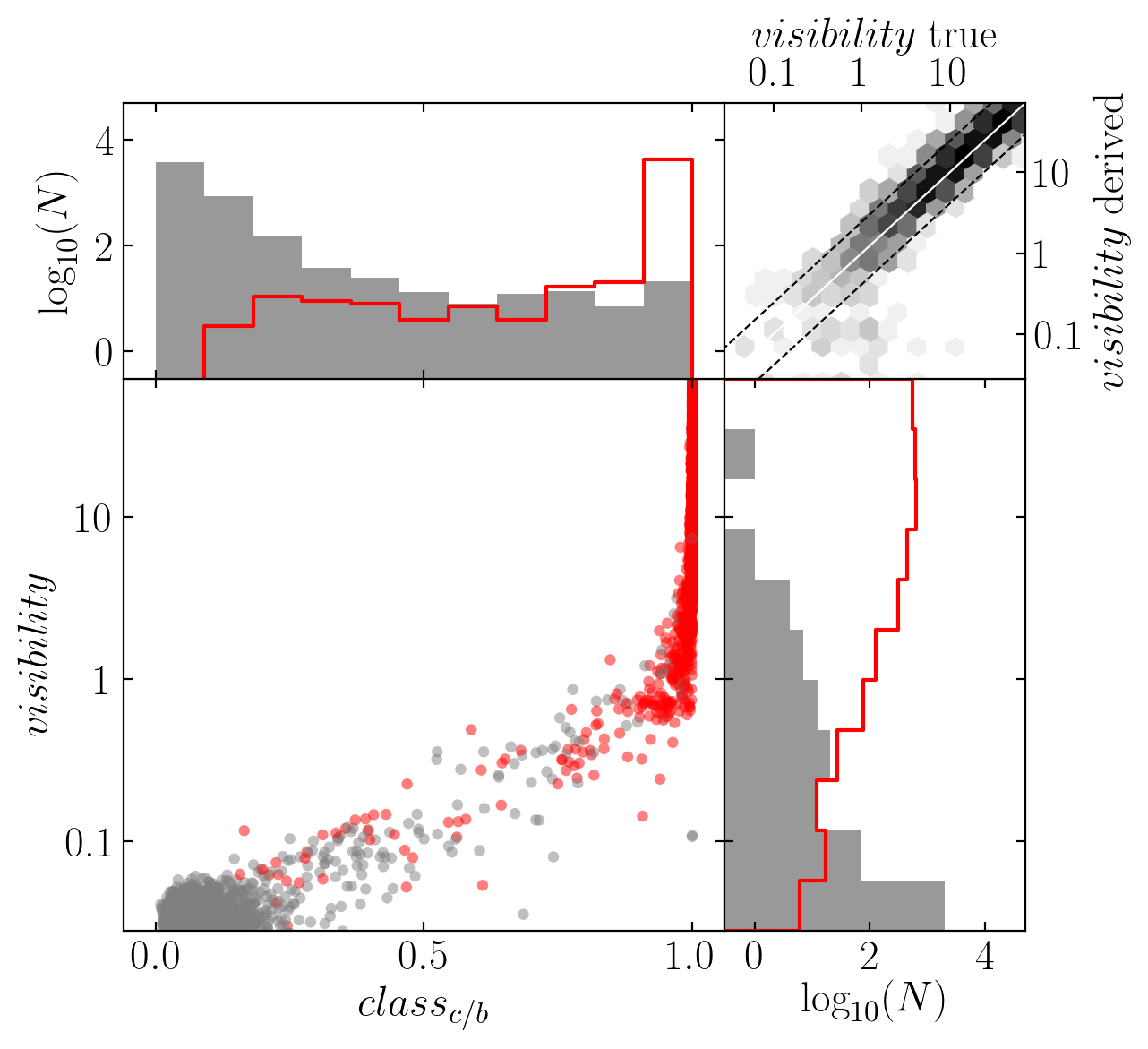}
    \else
        \includegraphics[width=1\columnwidth]{visibility_vs_clustericity.png}
    \fi
    \caption{$Visibility$ parameter values vs. $class_{c/b}$ for mock (red) and random M83 field (gray) samples. The histograms are marginalized logarithmic counts of samples for $visibility$ (right) and $class_{c/b}$ (top). Top-right panel shows true vs. derived $visibility$ parameters of mock clusters as a logarithmic density map; dashed lines outline the area containing 95\% of the clusters.}
    \label{fig:visibility_vs_clustericity}
\end{figure}

\begin{figure*}
    \centering
    \ifreferee
        \includegraphics[width=0.95\textwidth]{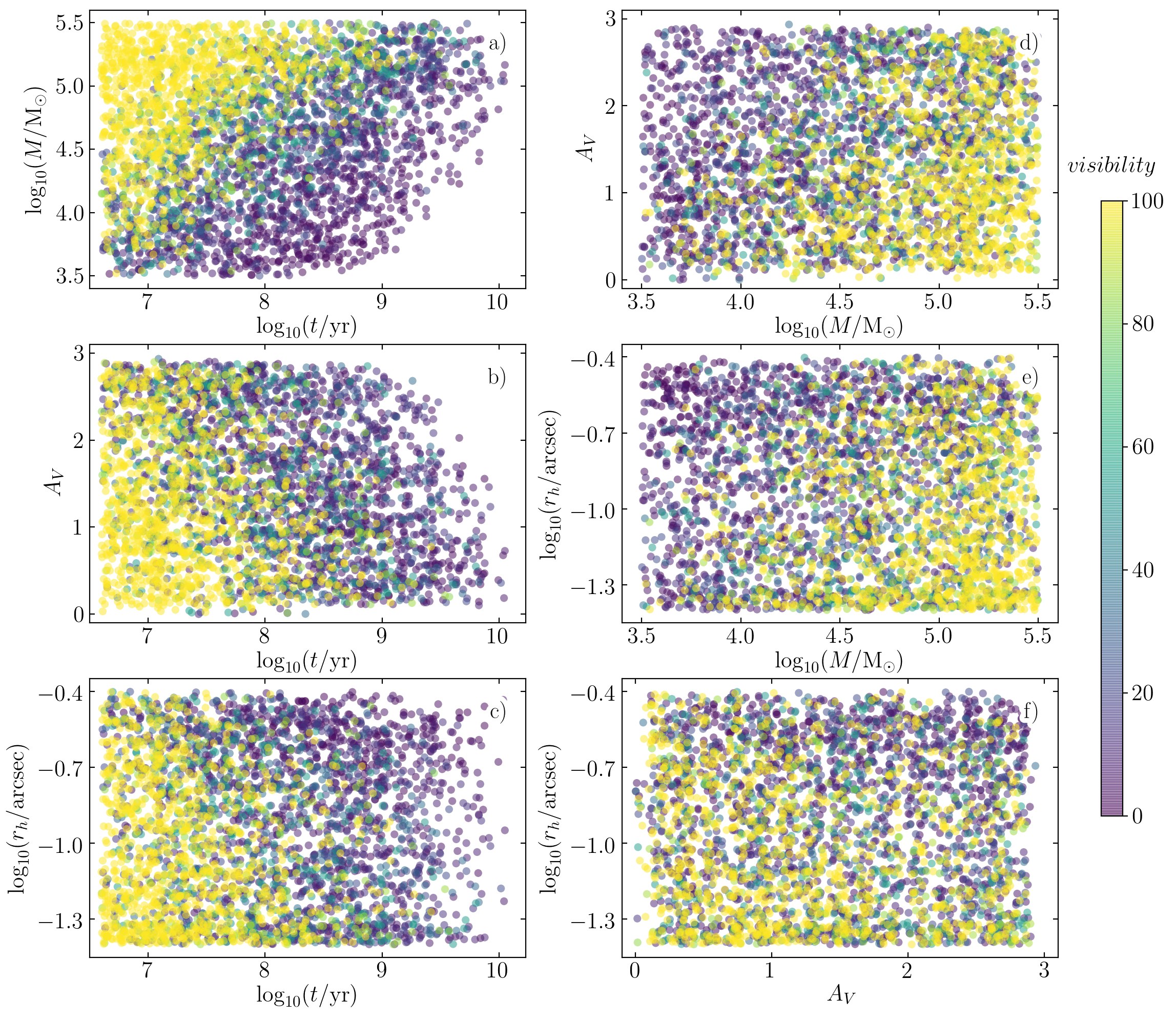}
    \else
        \includegraphics[width=0.85\textwidth]{param_param_visibility.jpg}
    \fi
    \caption{Inferred parameter distributions for the test cluster sample. The panels show the following parameter combinations: a) mass vs. age, b) extinction vs. age, c) size vs. age, d) extinction vs. mass, e) size vs mass, and f) size vs. extinction. Diagonal cutoffs in panels a) and b) are related to the cluster detection limit, applied as magnitude cuts, shown in Fig. \ref{fig:mock_cmd}. The color map represents the $visibility$ parameter, which acts as a proxy for the selection effects in a magnitude limited sample, while also taking into account variable cluster sizes and extinctions.}
    \label{fig:param_param_visibility}
\end{figure*}

\begin{figure*}
    \centering
    \ifreferee
        \includegraphics[width=0.9\textwidth]{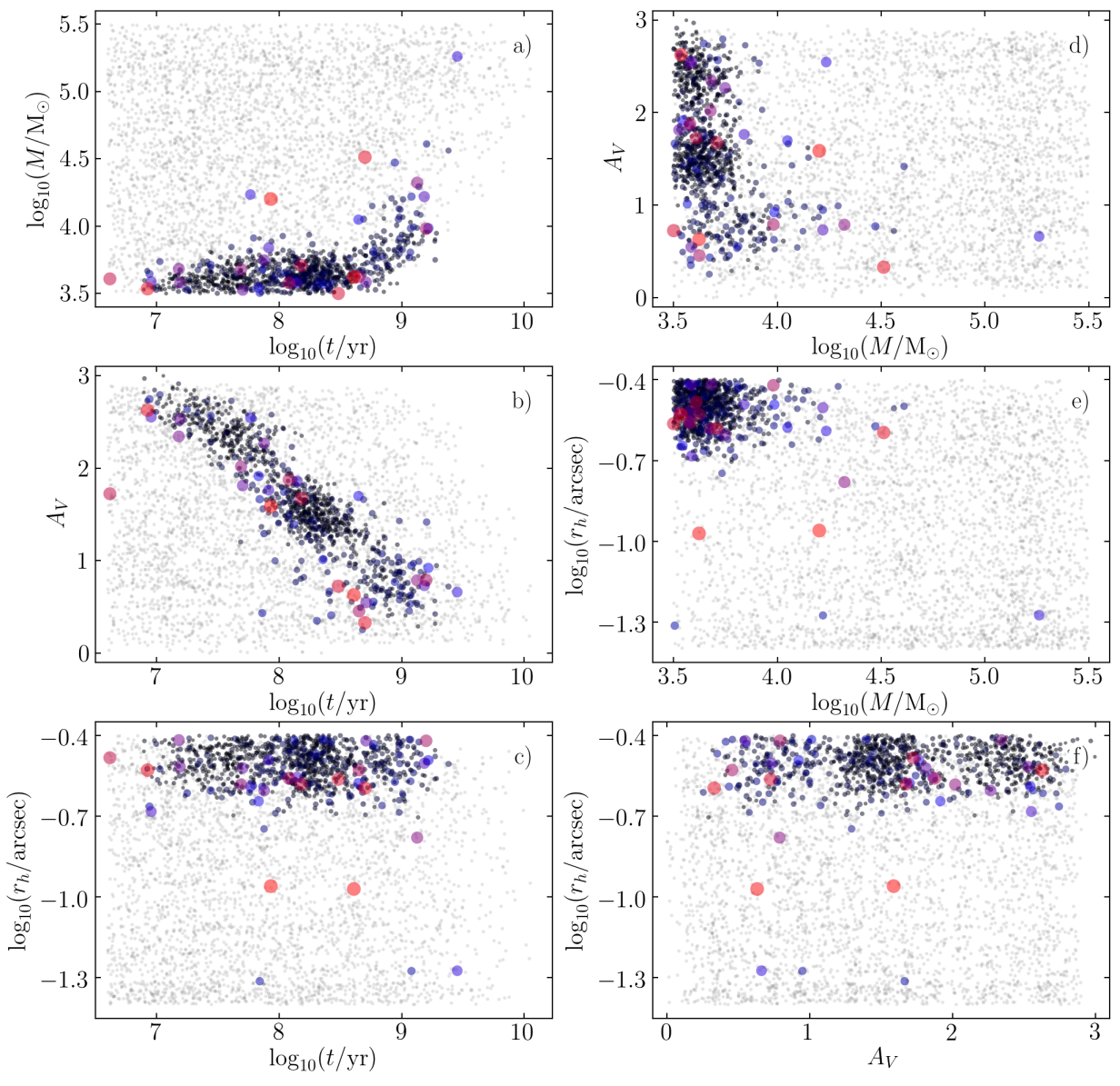}
    \else
        \includegraphics[width=0.75\textwidth]{param_param_bg.png}
    \fi
    \caption{Same as in Fig. \ref{fig:param_param_visibility}, but for 5,000 randomly spatially sampled M83 background images. For reference, derived mock cluster parameters are shown as faint gray dots. Real background samples are shown as black dots, that transition to blue and then to red. The dot size and color represents $class_{c/b}$. The blue dots are objects with $class_{c/b}>0.5$ and the red dots are objects with $class_{c/b}>0.99$.}
    \label{fig:param_param_bg}
\end{figure*}

Fig. \ref{fig:mock_diagonal_results}c shows the true and derived $A_V$ values. The 95\% of all inference results deviate $<$1.4 dex from the true values, as shown by the dashed lines. The highlighted blue and red clusters are classified as having significantly higher and lower extinction respectively. This can be explained by the age-extinction degeneracy, as older clusters with low extinction are hard to distinguish from younger clusters with high extinction, and vice-versa, when using only three photometric passbands.

In Fig. \ref{fig:mock_cmd}a-b the age-extinction degeneracy can be seen in the lower S-shaped part ($m_{F336W}-m_{F438W}>0$) of the color-color distribution of clusters. Clusters older than $\log_{10}(t/{\rm yr})=8$ with high extinction can be located in the same color-color area as clusters with low extinction. These effects have also been observed when using analytically integrated stellar luminosities \citep{2008BaltA..17..337B} and remain when stochastic effects of IMF sampling are included \citep{2014A&A...569A...4D}.

Fig. \ref{fig:mock_diagonal_results}b shows the true and derived mass values. Overall no systematic effects can be seen. The 95\% of all inference results deviate $<$0.4 dex from the true values, as shown by the dashed lines.

Fig. \ref{fig:mock_diagonal_results}d shows the true and derived $r_h$ values. No systematic effects can be seen. The 95\% of all inference results deviate $<$0.2 dex from the true values, as shown by the dashed lines. However, for the smallest clusters the error spread is as low as $\sim$0.1 dex, while for the largest clusters the error spread goes up to $\sim$0.2 dex. This can be explained by the clusters with higher $r_h$ having lower signal-to-noise, as their stars are spread out over a larger area in space.

Although in Fig. \ref{fig:mock_diagonal_results}b due to the age-extinction degeneracy we observe underestimated and overestimated cluster masses, size errors shown in panel d show no such bias. This can be explained by mass being a function of a cluster's magnitude as can be seen in Fig. \ref{fig:mock_cmd}g, which makes the network mispredict its value if age and extinction are also mispredicted. However, size has no impact on cluster magnitude or color.

As we use images normalized in a passband-independent manner, the influence of calibration accuracy to our method was also explored. Fig. \ref{fig:mock_diagonal_results_varbg} shows results obtained on the same dataset as Fig. \ref{fig:mock_diagonal_results}, only with the CNN trained on images with background fluxes that were varied from image to image. The flux scaling factor was sampled independently for each passband as a Gaussian with a mean of 1 and a standard deviation of 0.2. After multiplying the background image flux by this factor the cluster images were added and the final images normalized as usual. This encourages the network to learn parameter inference regardless of whether the calibrations for backgrounds match mock clusters well. As can be seen when comparing Figs. \ref{fig:mock_diagonal_results} and \ref{fig:mock_diagonal_results_varbg}, the inference results are very similar, only with the error spread increasing for each parameter by about 10\%. This implies that accurate calibrations, while still associated with slightly more precise results, are not essential for a CNN to derive cluster parameters.

Fig. \ref{fig:visibility_vs_clustericity} shows the derived $class_{c/b}$ and $visibility$ values for the 5,000 test mock clusters, as well as a random sample of 5,000 M83 background images. As can be seen in the histogram on top, the $class_{c/b}$ parameter is predicted as $>0.5$ for the vast majority of mock cluster images, and as $<0.5$ for the majority of background images. This suggests that the fraction of background images that are classified as $class_{c/b}>0.5$ are likely to correspond to real clusters. The $visibility$ parameter is highly correlated with $class_{c/b}$, again showing high values for the majority of mock clusters and low values for the majority of backgrounds. The few remaining mock clusters with $class_{c/b}<0.5$ have very low $visibility$ values, which indicates faint, nearly invisible objects seen in Fig. \ref{fig:artificial_cluster_samples_ageav}.

Fig. \ref{fig:param_param_visibility} illustrates selection effects by showing the derived age, extinction, mass, and size parameters of the test mock clusters, with the color bar representing the derived $visibility$ parameter value for each cluster. In Fig. \ref{fig:param_param_visibility}a it can be seen that mass and age are correlated as expected when deriving the $visibility$ parameter: clusters with lower mass and older ages tend to be less visible (this can also be seen in Figs. \ref{fig:artificial_cluster_samples_noextinction} and \ref{fig:artificial_cluster_samples_extinction}). The same is true for extinction (panels b and d), as higher extinctions tend to make cluster less visible, and size (panels c, d, and f), as more concentrated clusters stand out relative to their backgrounds.

Even though the cluster-related parameter inference results for background images have no inherit meaning, the CNN produces values for all of its output neurons regardless. Looking at these values can provide us with additional insights. For example, we would expect backgrounds to be classified as low-mass extended objects. Fig. \ref{fig:param_param_bg} shows the derived parameters for the background images from Fig. \ref{fig:visibility_vs_clustericity}, with dot size and color indicating $class_{c/b}$. Black dots are images with $class_{c/b}$ close to 0, while red circles are images with $class_{c/b}$ close to 1. As can be seen in Fig. \ref{fig:param_param_bg}e, the vast majority of the backgrounds are classified as low-mass extended objects as expected, with some probable cluster images being spread out more evenly through the parameter space. The derived age values of these images are spread out through the whole age range (panels a, b, and c), however extinctions are heavily correlated with ages as seen in panel b. As the network is trained to predict extinction and age values regardless of what the cluster's background looks like, there is no intuitive value that should be predicted for background images in this case. In effect the CNN avoids areas of age-extinction parameter space where the appearance of an observed object is either extremely blue (high-extinction high-age) or extremely red (low-extinction low-age), which can only be associated with genuine clusters, resulting in this diagonal effect.

The $class_{c/b}$ parameter was shown to be usable in differentiating between cluster and background images, while the $visibility$ parameter is correlated well with those cluster parameter ranges, which can show more confidently identified clusters. We conclude that these parameters can be useful indicators in star cluster search application.

\subsection{Validation with cataloged clusters}

\begin{figure}
    \centering
    \ifreferee
        \includegraphics[width=0.65\columnwidth]{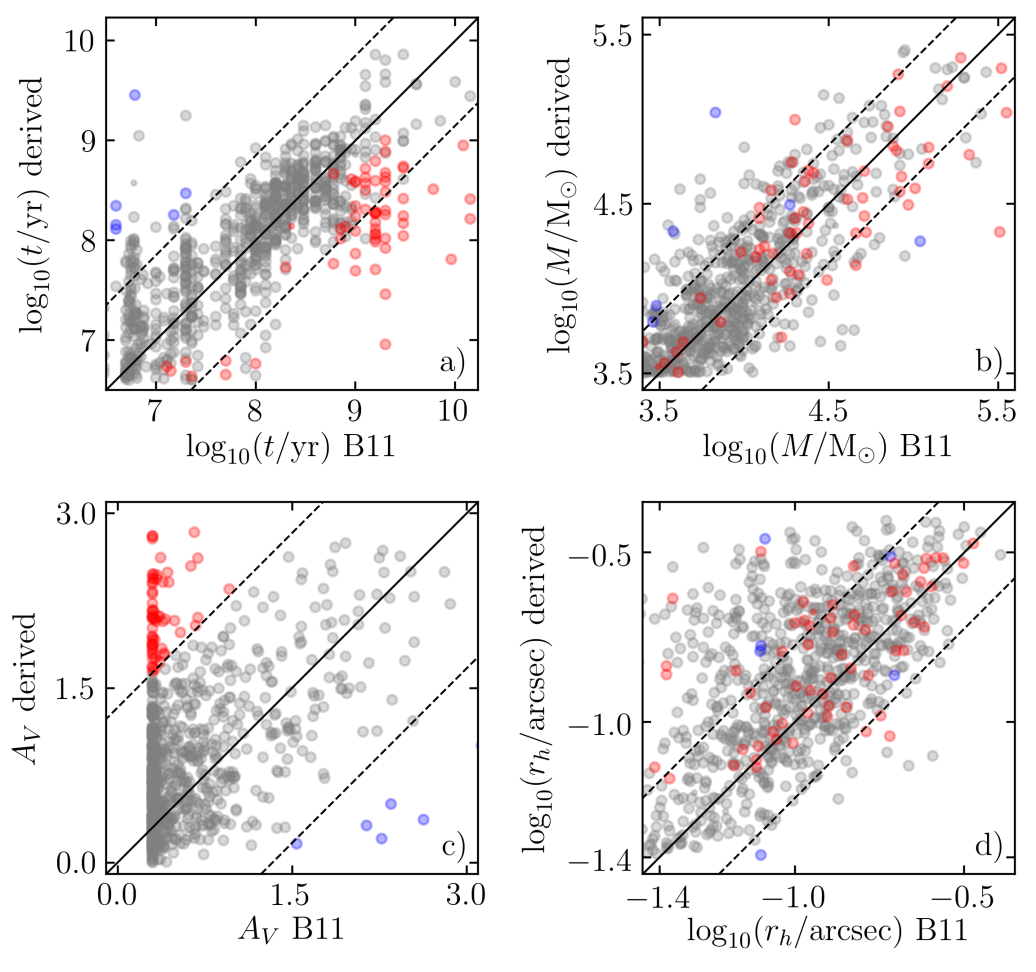}
    \else
        \includegraphics[width=1\columnwidth]{diagonal_real_differences_extinction.png}
    \fi
    \caption{The comparisons for a) age, b) mass, c) $A_V$, and d) $r_h$ derived by \cite{2011MNRAS.417L...6B} and by CNN. Red dots represent clusters with overestimated, while blue dots represent clusters with underestimated extinction values with respect to \cite{2011MNRAS.417L...6B}. The remaining clusters are colored gray. Dashed lines outline the area containing 95\% of the mock clusters.}
    \label{fig:bastian_diagonal_results_extinction}
\end{figure}

To validate our method on real clusters we used three previous M83 HST star cluster studies which had published catalogs. This includes the study covering the whole galactic disk (7 WFC3 fields) by \citet[R15]{2015MNRAS.452..525R}, two WFC3 fields by \citet[B11]{2011MNRAS.417L...6B} and the galaxies central region by \citet[H01]{Harris_2001}.

The study by \cite{2011MNRAS.417L...6B} is comprised of 939 objects. We discarded objects with missing parameter values, leaving us with 889 of them to compare to the CNN inference results. \cite{2011MNRAS.417L...6B} estimated the cluster age, mass, and extinction by comparing the integral photometry of the observed clusters to SSP models. Meanwhile, the sizes of clusters were estimated by fitting spatial models to F438W, F555W, and F814W band images. For this comparison we took the median value of these three size estimates. As the cluster magnitudes used by \cite{2011MNRAS.417L...6B} were Galactic extinction corrected, we shift the $A_V$ values of those objects by 0.3 mag\footnote{https://irsa.ipac.caltech.edu/applications/DUST/}. This was done so that we could compare CNN derived values directly, because we compute total extinctions for clusters regardless of the dust source.

\begin{figure}
    \centering
    \ifreferee
        \includegraphics[width=0.65\columnwidth]{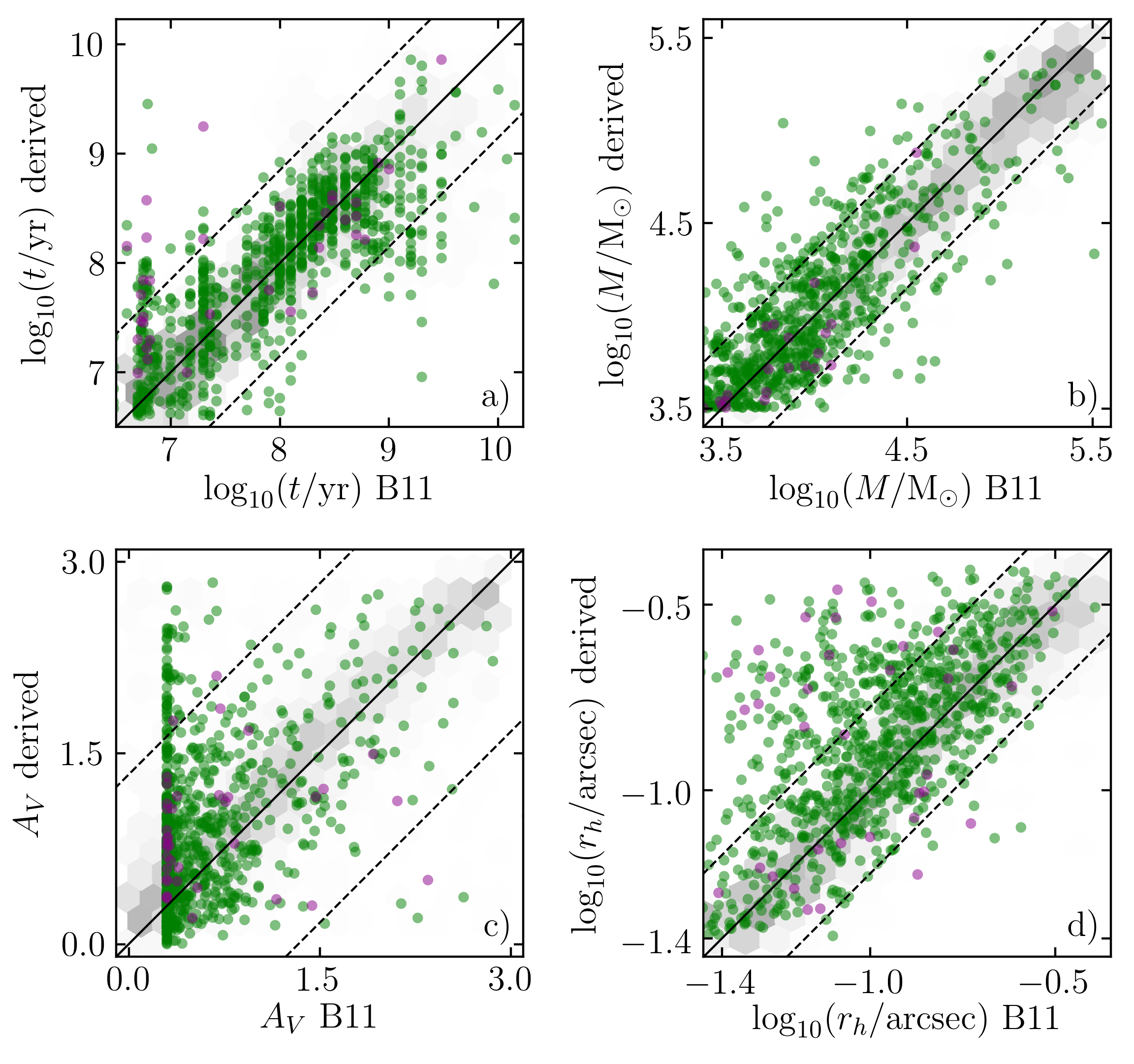}
    \else
        \includegraphics[width=1\columnwidth]{diagonal_real_differences.png}
    \fi
    \caption{Same as Fig. \ref{fig:bastian_diagonal_results_extinction}, but with large green dots representing objects classified as likely clusters ($class_{c/b}\geq0.5$), and magenta dots representing objects with $class_{c/b}<0.5$. Mock clusters are displayed in the background as a hexagonal density maps with bins scaled linearly.}
    \label{fig:bastian_diagonal_results}
\end{figure}

\begin{figure}
    \centering
    \ifreferee
        \includegraphics[width=0.65\columnwidth]{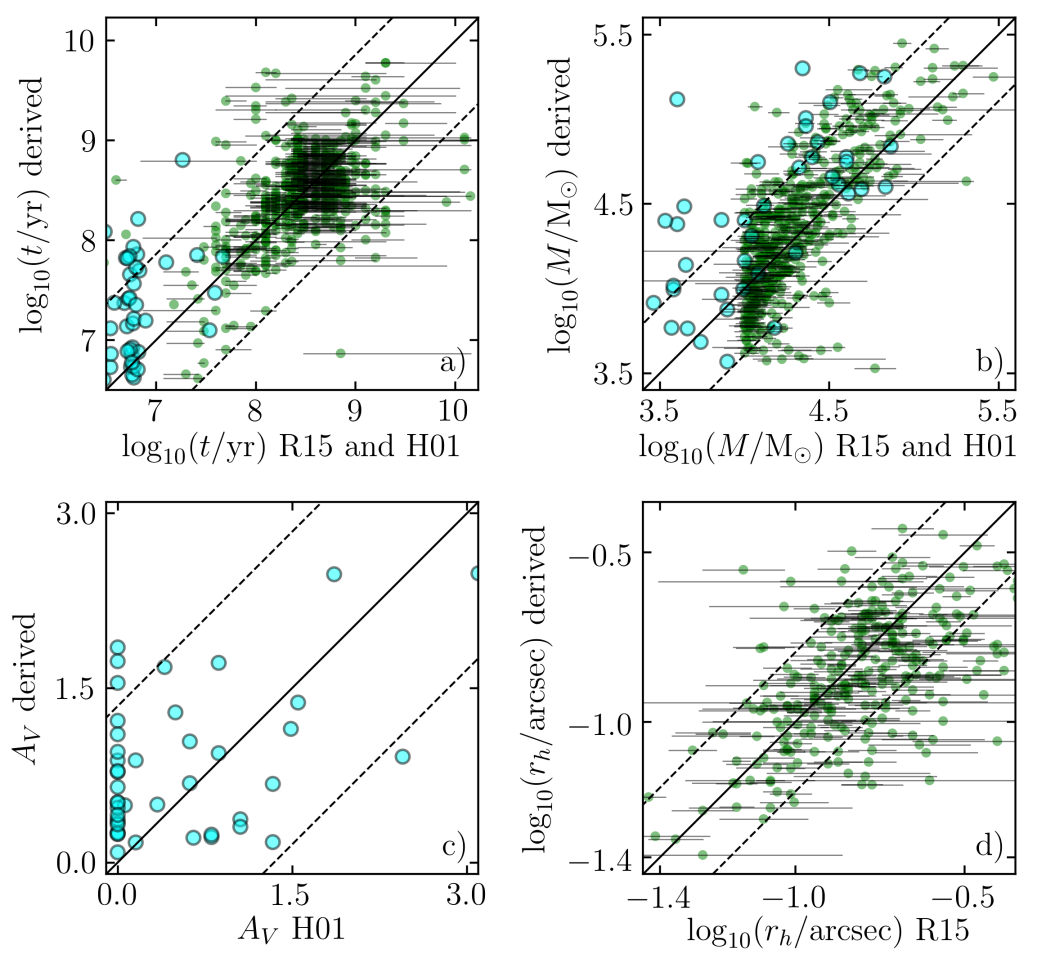}
    \else
        \includegraphics[width=1\columnwidth]{diagonal_ryonharris_differences.png}
    \fi
    \caption{Same as Fig. \ref{fig:bastian_diagonal_results}, but with green dots representing objects from \cite{2015MNRAS.452..525R} and cyan dots representing objects from \cite{Harris_2001}. The horizontal bars denote minimum and maximum parameter values for age and mass, and statistical errors for size, as provided in the catalogs.}
    \label{fig:ryonharris_diagonal_results}
\end{figure}

\begin{figure*}
    \centering
    \ifreferee
        \includegraphics[width=0.9\textwidth]{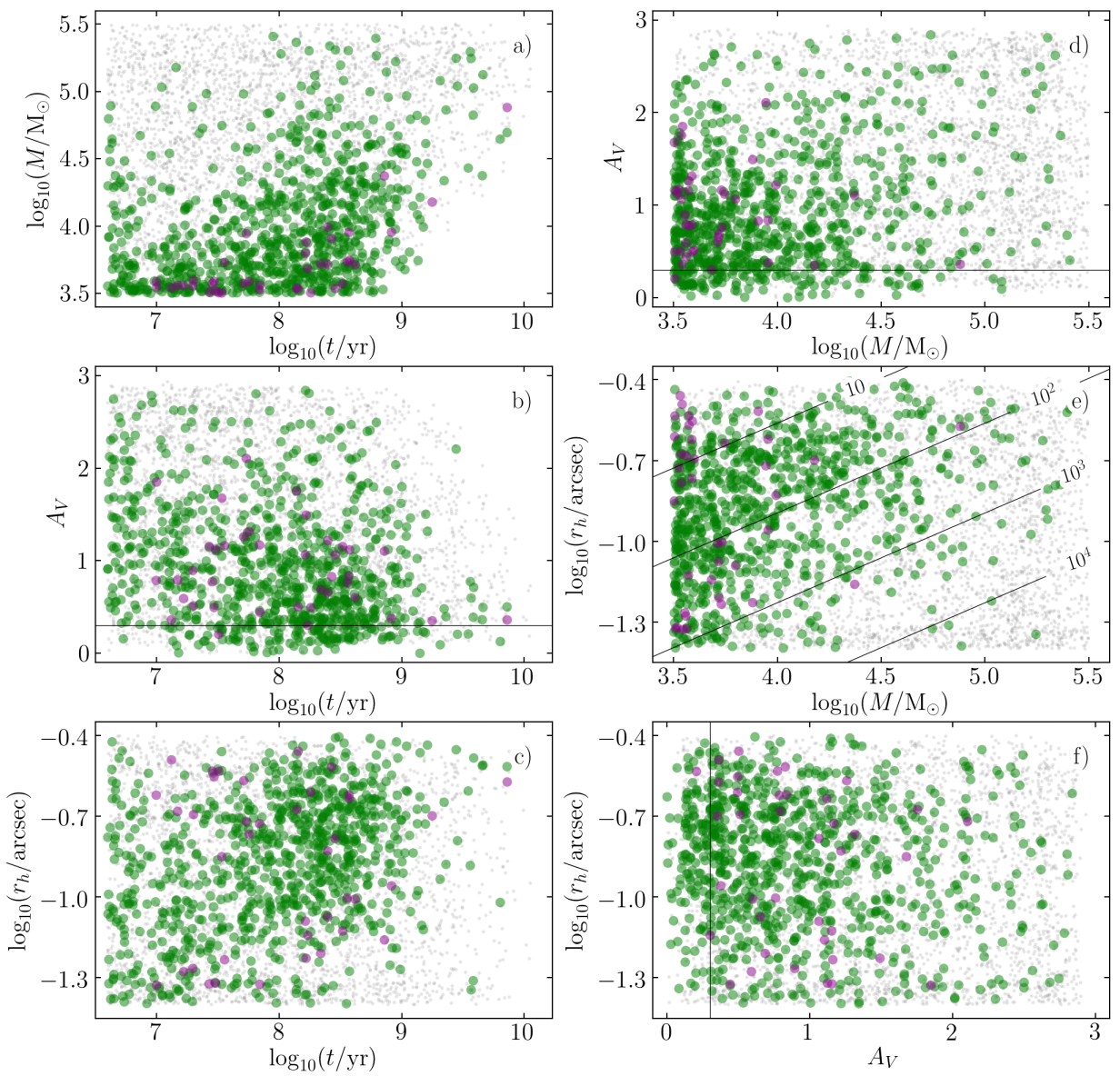}
    \else
        \includegraphics[width=0.75\textwidth]{param_param_real.png}
    \fi
    \caption{Same as Fig. \ref{fig:param_param_visibility}, but for objects from \cite{2011MNRAS.417L...6B}, with values derived by CNN. The green circles represent objects classified as likely clusters, while magenta circles represent likely non-clusters, as in Fig. \ref{fig:bastian_diagonal_results}. For reference, the derived parameters of the mock cluster set are shown as faint gray points. In panel e lines show locations of clusters with the same density, varying from 10 to $10^4$ ${\rm M}_\odot / {\rm pc}^{-3}$. In panels b, d, and f the solid black lines represent the amount of Galactic extinction in the direction of M83 ($A_V=0.3$ mag).}
    \label{fig:param_param_real}
\end{figure*}

Figs. \ref{fig:bastian_diagonal_results_extinction} and \ref{fig:bastian_diagonal_results} show a comparison between \cite{2011MNRAS.417L...6B} and CNN-derived values. In Fig. \ref{fig:bastian_diagonal_results_extinction} the red and blue dots represent clusters with significantly overestimated and underestimated extinction values respectively. They were defined as clusters that are outside the dashed lines in panel c, which represent the area containing 95\% of mock cluster parameter derivations. This mirrors the situation with mock objects in Fig. \ref{fig:mock_diagonal_results}, as the majority of clusters with overestimated extinctions end up with underestimated ages, and vice-versa for clusters with underestimated extinction values. These effects can again be attributed to the age-extinction degeneracy. In Fig. \ref{fig:bastian_diagonal_results} the green dots represent images classified by the network as likely to be real clusters ($class_{c/b}\geq0.5$), while the magenta dots are objects with $class_{c/b}<0.5$. The vast majority of the objects are classified as likely clusters.

Overall the derived ages and masses show a reasonable correlation between \cite{2011MNRAS.417L...6B} and CNN-derived values. Many of the objects have cataloged $A_V=0$ mag values (shown as $A_V=0.3$ mag in the figures, accounting for Galactic extinction). The CNN derives higher extinctions for some of these clusters, however, visual inspection has revealed that Galactic dust is unlikely to be the only source of extinction for the majority of them. The sizes show a good agreement for most of the objects, however, there is a subset of objects with somewhat overestimated values.

For the comparison with \cite{2015MNRAS.452..525R} we used 478 objects which had sizes obtained by a 2D spatial model fitting as well as age and mass estimates derived using spectral energy distribution fitting. We also took 45 objects from \cite{Harris_2001} with their age, mass, and extinction estimates obtained by comparing the cluster photometry to theoretical population synthesis models. Fig. \ref{fig:ryonharris_diagonal_results} shows our results compared against both of these catalogs. \cite{2015MNRAS.452..525R} objects are denoted as green dots with the parameter error bounds marked with black lines. \cite{Harris_2001} objects are marked as large cyan circles. For both of these catalogs a reasonable agreement with the CNN-derived values can be seen, with only masses being slightly overestimated. However, there's some age estimate divergence over $\log_{10}(t/{\rm yr})=8$, which is similar to the situation in Fig. \ref{fig:bastian_diagonal_results_extinction}.

\begin{figure*}
    \centering
        \includegraphics[width=1.0\textwidth]{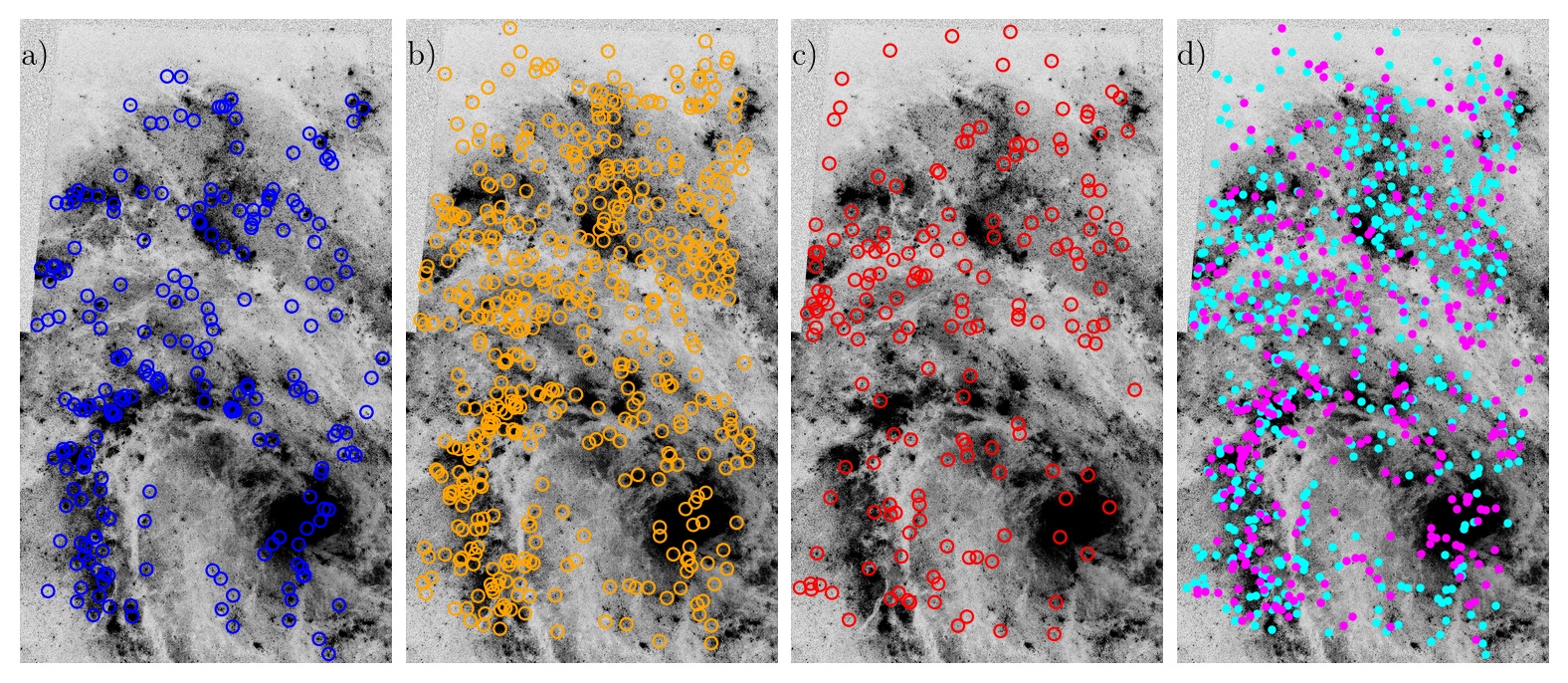}
    \caption{M83 mozaic of the F438W passband observations, overplotted with young -- $\log_{10}(t/{\rm yr})<7.4$ (blue), intermediate -- $7.4\geq\log_{10}(t/{\rm yr})<8.6$ (yellow), and old -- $\log_{10}(t/{\rm yr})\geq8.6$ (red) objects from \cite{2011MNRAS.417L...6B}. The last panel depicts object with low -- $A_V<1$ mag (cyan) and high -- $A_V\geq1$ mag (magenta) extinctions.}
    \label{fig:m83_clusters}
\end{figure*}

\begin{figure}
    \centering
    \ifreferee
        \includegraphics[width=0.6\columnwidth]{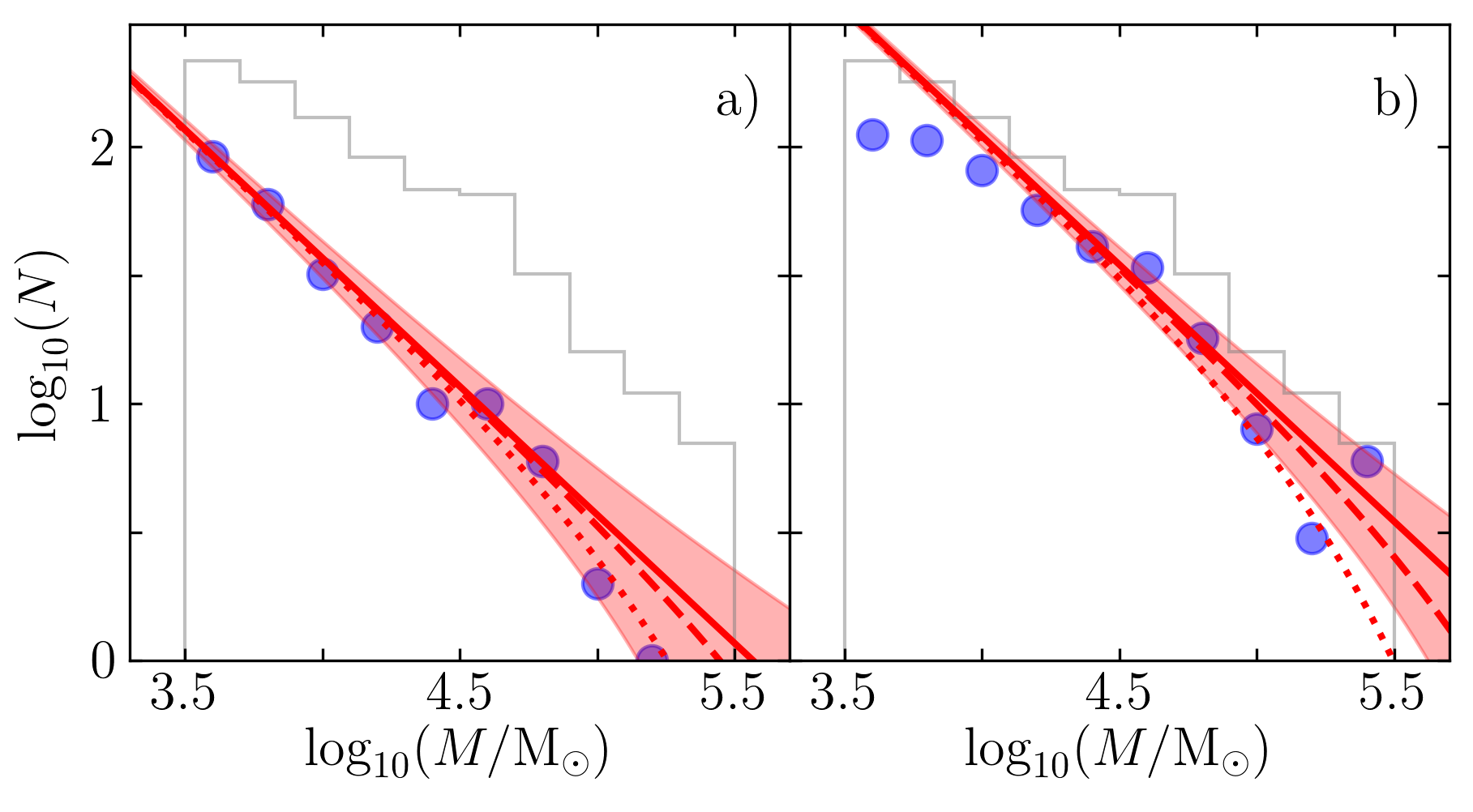}
    \else
        \includegraphics[width=0.95\columnwidth]{age_mass_hists_twoslice.png}
    \fi
    \caption{Cluster mass distributions for samples with ages $\log_{10}(t/{\rm yr})<7.7$ (a) and $7.7<\log_{10}(t/{\rm yr})<8.7$ (b). Lines represent the power law mass distribution function of the form $dN / dM = A \cdot M^{-2} \cdot \exp(-M / M_*)$: $M_* = \infty$ (solid line, with the shaded area encompassing its Poisson standard deviation), $10^6$ $M_\odot$ (dashed line), and $2.5 \cdot 10^5$ $M_\odot$ (dotted line).}
    \label{fig:age_mass_hists_twoslice}
\end{figure}

\begin{figure*}
    \centering
    \includegraphics[width=0.9\textwidth]{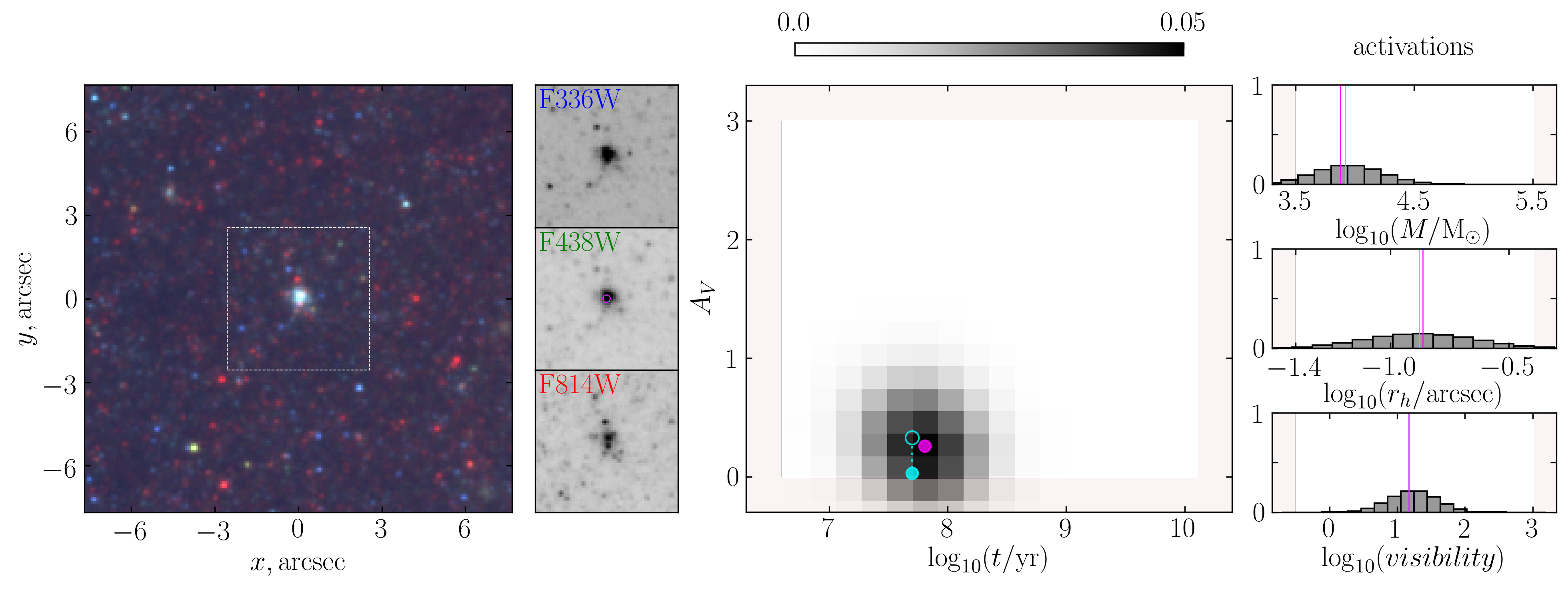}
    \includegraphics[width=0.9\textwidth]{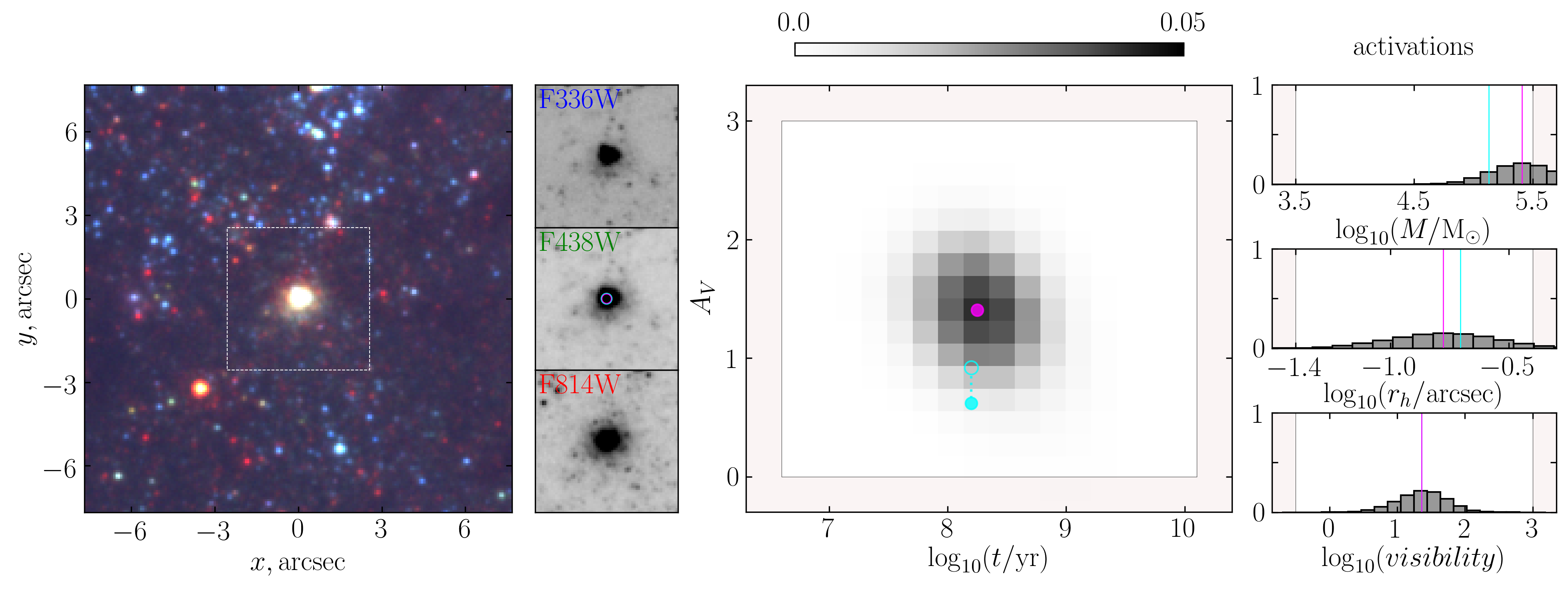}
    \includegraphics[width=0.9\textwidth]{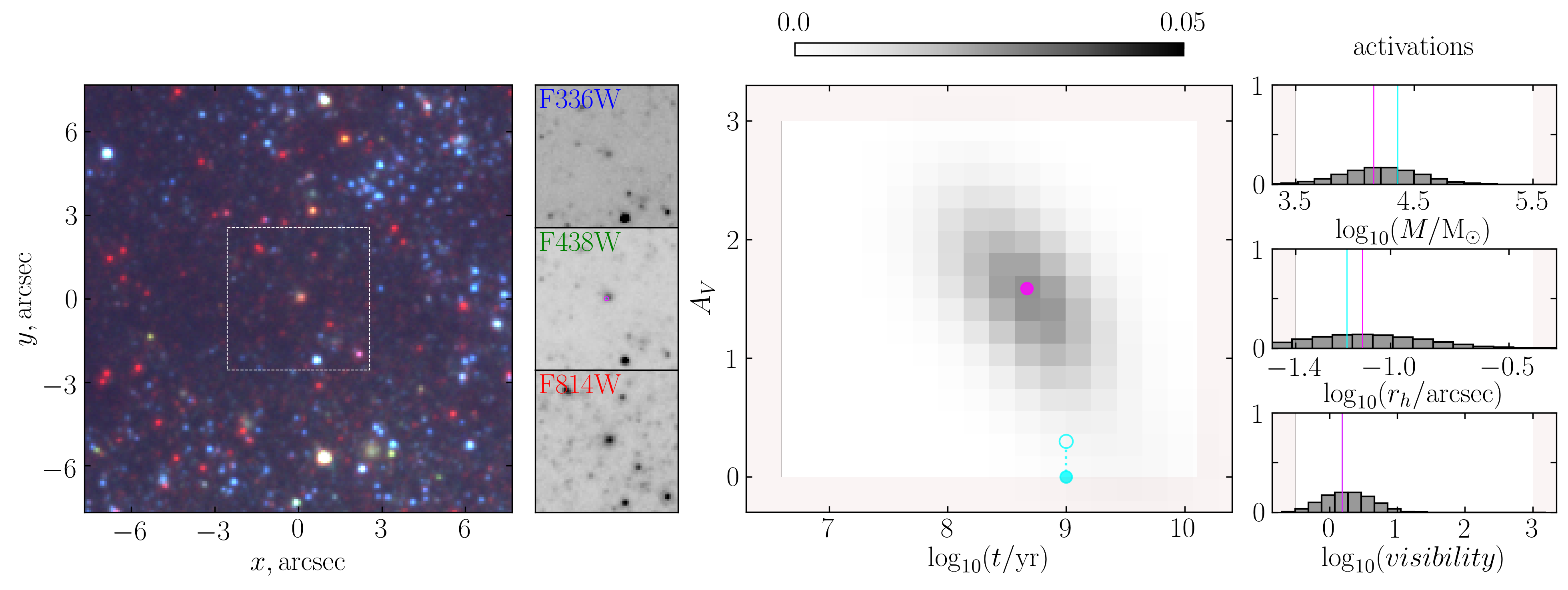}
    \caption{Examples of real clusters and their parameter distributions inferred by the CNN. The left panels show the color image of each cluster in a field of $7\arcsec\times7\arcsec$, with a $2.6\arcsec\times2.6\arcsec$ field used for inference highlighted with a dashed square. The three grayscale images are the three passband observations shown separately. All of the images are normalized as in Fig. \ref{fig:artificial_cluster_samples_noextinction}. The remaining panels depict inference results, with the histograms showing the neural network's neuron activations for the given parameter. Age and extinction is depicted as a 2D activation map marginalized over mass to highlight the effects of age-extinction degeneracies, with the color bar on the top indicating CNN output neuron activation strength. The \cite{2011MNRAS.417L...6B} derived values of parameters are marked in cyan, and the CNN inferred values in magenta. Light-red shaded areas show parameter ranges where the CNN produces activations, but which were not covered by the clusters used in this study to deal with parameter boundary effects. The empty cyan circle in the age-extinction map represents the values obtained after Galactic extinction correction.}
    \label{fig:example_inference_results}
\end{figure*}

We have shown that the CNN is capable of deriving cluster parameters on real clusters by comparing our results with those of other authors. The agreements between the values are reasonable and follow the results obtained with mock clusters. However, due to the age-extinction degeneracy with the used 3 passbands, the results with clusters older than $\log_{10}(t/{\rm yr})=8$ are ambiguous and should be interpreted carefully.

\section{Discussion}
\label{sec:discussion}

We have shown the applicability of a CNN-based method in deriving a variety of star cluster parameters from M83 mosaic images in terms of quantitative error analysis. However, the final aim for this method is to be of use in star cluster search and automatic catalog construction. To this end, a better look into the derived parameters is needed both in terms of each other, and their context in the galaxy. In this chapter we look at derived values of the \cite{2011MNRAS.417L...6B} sample of objects in more detail.

Fig. \ref{fig:param_param_real} shows the inferred age, extinction, mass, and size parameters of the \cite{2011MNRAS.417L...6B} object sample. The objects are colored as in Fig. \ref{fig:bastian_diagonal_results}, with mock results shown in the background. The clusters cover the whole parameter range well, with $class_{c/b}<0.5$ samples being classified as expected: as low-mass objects (panel a). The minimal extinction line, with a large number of clusters around it, seen in panels b, d, and f, coincides with $A_V\sim0.3$ mag, expected due to Galactic dust foreground in the direction of M83. Lines of constant density are shown in panel e. The majority of the objects fall within 10 and $10^3$ ${\rm M}_\odot / {\rm pc}^{-3}$, which is consistent with results for clusters of the M31 galaxy \citep{2009ApJ...703.1872V}.

Fig. \ref{fig:m83_clusters} shows \cite{2011MNRAS.417L...6B} objects marked on two fields of the M83 mosaic. Objects of $\log_{10}(t/{\rm yr})<7.4$ are marked as blue circles in panel a, $\log_{10}(t/{\rm yr})<8.6$ objects are marked as orange circles in panel b, and $\log_{10}(t/{\rm yr})\geq8.6$ objects are marked as red circles in panel c. Panel d shows all of the objects marked as dots, with $A_V<1$ mag colored cyan, and $A_V\geq1$ mag colored magenta. The spatial distribution of objects is sensible, with young star clusters grouping around the galaxies spiral arms, near the dust clouds where they were formed, and old clusters spread out more evenly throughout the galaxy, as they had more time to drift away. The extinction distributions are less clear-cut, however some crowding around dust-heavy regions can be seen by the high-extinction objects, as is expected. The spatial distributions of age-selected clusters in Fig. \ref{fig:m83_clusters} correspond well to the results obtained by \cite{2012ApJ...750...60F} using UBVIH$\alpha$ fluxes to measure ages, masses, and extinctions in the central region of M83. \cite{2019MNRAS.483.2641S} has derived age maps for the M83 galaxy's stellar populations younger than 20 Myr, which corresponds to the lower age range of clusters in this study.

Although we studied clusters with masses $\log(M/{\rm M_\odot})=[3.5, 5.5]$, this does not imply that only such clusters are detectable with the HST/WFC3 observations of M83. In fact, clusters with masses as low as $\log(M/{\rm M_\odot}) \sim 3$ have been studied by \cite{2011ApJ...729...78W} and \cite{2014ApJ...793....4A}. However, such clusters are dominated by stochastic effects of IMF sampling making the analysis of the effects of extinction problematic. The lower-limit of masses was selected to focus on the effects of extinction as well as to align with the range of clusters used by \cite{2011MNRAS.417L...6B}. The presented CNN classifies lower mass clusters as being on the lower-limit of this range.

Fig. \ref{fig:age_mass_hists_twoslice} shows the binned mass distributions obtained with the CNN. The gray outline shows all of the cluster distributions, with blue dots representing clusters of $\log_{10}(t/{\rm yr})<7.7$ (panel a), and $7.7\leq\log_{10}(t/{\rm yr})<8.7$ (panel b). The red lines represent Schechter type mass functions \citep{2010ARA&A..48..431P} with various amounts of truncation. The solid red line follows the non-truncated power law $dN / dM = A \cdot M^{-2}$, the dashed red line follows $A \cdot M^{-2} \cdot \exp(\frac{-M}{10^6})$, and the dotted red line follows $A \cdot M^{-2} \cdot \exp(\frac{-M}{2.5 \cdot 10^5})$. The power law distributions fit the data well for both of the age cuts, however, there is a lack of low-mass clusters ($\log_{10}(M/{\rm M_\odot})\leq4$) for the mid-age data sample. This is due to selection effects, with less star clusters being detectable at those ages (see Fig. \ref{fig:param_param_real}a). Similar cluster mass distributions and selection effects have been found in M31 \citep{2009ApJ...703.1872V} and M33 \citep{2015A&A...581A.111D} star cluster samples.

Fig. \ref{fig:example_inference_results} show examples of inference results on 3 distinct \cite{2011MNRAS.417L...6B} clusters chosen to illustrate the variety of CNN outputs (previously sketched in Fig. \ref{fig:cnn_outputs}). The top row shows a young, low-mass cluster. The inferred age and mass matches \cite{2011MNRAS.417L...6B} parameters well. Extinction is derived to be slightly higher, however the value is very close when Galactic extinction is accounted for. The $visibility$ parameter is derived to be $\sim$15, which corresponds well to similarly looking clusters in Fig. \ref{fig:artificial_cluster_samples_extinction}b.

In the middle, a cluster of $\log_{10}(t/{\rm yr}) \sim 8.3$, medium extinction and high mass is shown. The age, mass, and size correspond well to the values derived by \cite{2011MNRAS.417L...6B}. Extinction is derived as slightly higher, however it's still within the range of CNN's activations. The cluster is classified as brighter by the CNN, with $visibility\sim25$.

On the bottom an older cluster is depicted. Its mass and size estimates correspond well, however extinction is overestimated in comparison to \cite{2011MNRAS.417L...6B}. Furthermore, the neuron activations show a diagonal pattern highlighting the age-extinction degeneracy which is hard to resolve with the used 3 passbands. However, the higher-extinction results are more likely as a significant amount of the field seen in the leftmost panel appears reddened, which suggest the presence of dust obscuring the cluster.

As detailed in Section \ref{sec:stochastic_effects}, a correlation between the spread of CNN output activations and scatter of inferred cluster parameters is noted, therefore, activation maps can be used to estimate cluster parameter uncertainties. We checked that less than 1\% of the clusters show the bimodal distribution activation distributions and about 20\% of the samples show an extended unimodal distribution (see Fig. \ref{fig:stochastic_effects} for examples). The rest of the results are unimodal. Therefore, selecting the highest activation and interpolating it is a viable approach to provide inferred parameter estimates. However, there are some cases where there is a systematic bias in the derived results. This usually occurs for the nearly invisible clusters as well as clusters with high extinctions and older ages, where age-extinction degeneracies make inference unreliable with the used photometric passbands. This means that the extent of activation maps alone, while informative, is not a reliable uncertainty estimate in all cases. Additional insights on the reliability of inference results can be gained by performing the random background sampling test described in Section \ref{sec:stochastic_effects}.

These results further validate the applicability of the CNN in deriving the parameters of star clusters in realistic scenarios. In addition, the $class_{c/b}$ and $visibility$ parameters act as accurate proxies for cluster presence in images. Utilizing this method for constructing a full catalog of M83 clusters is left for the subsequent paper in the series.

\section{Conclusions}

We have extended the method introduced in Paper I to infer cluster ages, masses, sizes, extinctions, as well as to account for the degeneracies between them. Additional parameters were added for identifying the presence of clusters on background images of M83, and judging their visibility (signal-to-noise).

To train this network a bank of mock clusters was generated utilizing three photometric passbands in the context of the M83 galaxy. The CNN was verified on mock images of artificial clusters with ages, $\log_{10}(t/{\rm yr})$, between 6.6 and 10.1, masses, $\log_{10}(M/{\rm M_\odot})$, between 3.5 and 5.5, sizes between 0.04 and 0.4 arcsec, and extinctions $A_V\leq3$ mag. Parameters derived by CNN have shown a good agreement with the true parameters for $\log_{10}(t/{\rm yr})<8$, with higher age estimates being unreliable due to the age-extinction degeneracy.

Real cluster parameter inference tests were performed with three different M83 cluster catalogs from \cite{2011MNRAS.417L...6B}, \cite{2015MNRAS.452..525R}, and \cite{Harris_2001} and have shown consistent results.

We have demonstrated that a CNN can perform evolutionary (age, mass), structural (size), and environmental (extinction) star cluster parameter inference. In addition, the network is capable of giving an indication of cluster presence in images. Therefore, the created CNN is a useful tool for further research in constructing a full pipeline of star cluster detection and parameter inference.

\begin{acknowledgements}
This research was funded by a grant (No. LAT-09/2016) from the Research Council of Lithuania.

This research made use of Astropy, a community-developed core Python package for Astronomy (Astropy Collaboration, 2018).

Some of the data presented in this paper were obtained from the Mikulski Archive for Space Telescopes (MAST). STScI is operated by the Association of Universities for Research in Astronomy, Inc., under NASA contract NAS5-26555.

We are thankful to the anonymous referee who helped improve the paper.
\end{acknowledgements}

\bibliography{library}

\begin{thebibliography}{34}
\expandafter\ifx\csname natexlab\endcsname\relax\def\natexlab#1{#1}\fi

\bibitem[{{Andrews} {et~al.}(2014){Andrews}, {Calzetti}, {Chandar},
  {Elmegreen}, {Kennicutt}, {Kim}, {Krumholz}, {Lee}, {McElwee}, {O'Connell},
  \& {Whitmore}}]{2014ApJ...793....4A}
{Andrews}, J.~E., {Calzetti}, D., {Chandar}, R., {et~al.} 2014, \apj, 793, 4

\bibitem[{{Bastian} {et~al.}(2011){Bastian}, {Adamo}, {Gieles}, {Lamers},
  {Larsen}, {Silva-Villa}, {Smith}, {Kotulla}, {Konstantopoulos}, {Trancho}, \&
  {Zackrisson}}]{2011MNRAS.417L...6B}
{Bastian}, N., {Adamo}, A., {Gieles}, M., {et~al.} 2011, \mnras, 417, L6

\bibitem[{{Bialopetravi{\v{c}}ius} {et~al.}(2019){Bialopetravi{\v{c}}ius},
  {Narbutis}, \& {Vansevi{\v{c}}ius}}]{PaperI}
{Bialopetravi{\v{c}}ius}, J., {Narbutis}, D., \& {Vansevi{\v{c}}ius}, V. 2019,
  \aap, 621, A103

\bibitem[{{Blair} {et~al.}(2014){Blair}, {Chandar}, {Dopita}, {Ghavamian},
  {Hammer}, {Kuntz}, {Long}, {Soria}, {Whitmore}, \&
  {Winkler}}]{2014ApJ...788...55B}
{Blair}, W.~P., {Chandar}, R., {Dopita}, M.~A., {et~al.} 2014, \apj, 788, 55

\bibitem[{{Bressan} {et~al.}(2012){Bressan}, {Marigo}, {Girardi}, {Salasnich},
  {Dal Cero}, {Rubele}, \& {Nanni}}]{2012MNRAS.427..127B}
{Bressan}, A., {Marigo}, P., {Girardi}, L., {et~al.} 2012, \mnras, 427, 127

\bibitem[{{Brid{\v z}ius} {et~al.}(2008){Brid{\v z}ius}, {Narbutis},
  {Stonkut{\.e}}, {Deveikis}, \& {Vansevi{\v c}ius}}]{2008BaltA..17..337B}
{Brid{\v z}ius}, A., {Narbutis}, D., {Stonkut{\.e}}, R., {Deveikis}, V., \&
  {Vansevi{\v c}ius}, V. 2008, Baltic Astronomy, 17, 337

\bibitem[{{de Meulenaer} {et~al.}(2013){de Meulenaer}, {Narbutis}, {Mineikis},
  \& {Vansevi{\v c}ius}}]{2013A&A...550A..20D}
{de Meulenaer}, P., {Narbutis}, D., {Mineikis}, T., \& {Vansevi{\v c}ius}, V.
  2013, \aap, 550, A20

\bibitem[{{de Meulenaer} {et~al.}(2014){de Meulenaer}, {Narbutis}, {Mineikis},
  \& {Vansevi{\v c}ius}}]{2014A&A...569A...4D}
{de Meulenaer}, P., {Narbutis}, D., {Mineikis}, T., \& {Vansevi{\v c}ius}, V.
  2014, \aap, 569, A4

\bibitem[{{de Meulenaer} {et~al.}(2015){de Meulenaer}, {Narbutis}, {Mineikis},
  \& {Vansevi{\v{c}}ius}}]{2015A&A...581A.111D}
{de Meulenaer}, P., {Narbutis}, D., {Mineikis}, T., \& {Vansevi{\v{c}}ius}, V.
  2015, \aap, 581, A111

\bibitem[{{Dieleman} {et~al.}(2015){Dieleman}, {Willett}, \&
  {Dambre}}]{dieleman}
{Dieleman}, S., {Willett}, K.~W., \& {Dambre}, J. 2015, \mnras, 450, 1441

\bibitem[{{Dopita} {et~al.}(2010){Dopita}, {Blair}, {Long}, {Mutchler},
  {Whitmore}, {Kuntz}, {Balick}, {Bond}, {Calzetti}, {Carollo}, {Disney},
  {Frogel}, {O'Connell}, {Hall}, {Holtzman}, {Kimble}, {MacKenty}, {McCarthy},
  {Paresce}, {Saha}, {Silk}, {Sirianni}, {Trauger}, {Walker}, {Windhorst}, \&
  {Young}}]{2010ApJ...710..964D}
{Dopita}, M.~A., {Blair}, W.~P., {Long}, K.~S., {et~al.} 2010, \apj, 710, 964

\bibitem[{{Dressel}(2012)}]{2012wfci.book.....D}
{Dressel}, L. 2012, {Wide Field Camera 3 Instrument Handbook for Cycle 21 v.
  5.0}

\bibitem[{{Elson} {et~al.}(1987){Elson}, {Fall}, \&
  {Freeman}}]{1987ApJ...323...54E}
{Elson}, R.~A.~W., {Fall}, S.~M., \& {Freeman}, K.~C. 1987, \apj, 323, 54

\bibitem[{{Fouesneau} \& {Lan{\c c}on}(2010)}]{2010A&A...521A..22F}
{Fouesneau}, M. \& {Lan{\c c}on}, A. 2010, \aap, 521, A22

\bibitem[{{Fouesneau} {et~al.}(2012){Fouesneau}, {Lan{\c{c}}on}, {Chandar}, \&
  {Whitmore}}]{2012ApJ...750...60F}
{Fouesneau}, M., {Lan{\c{c}}on}, A., {Chandar}, R., \& {Whitmore}, B.~C. 2012,
  \apj, 750, 60

\bibitem[{Harris {et~al.}(2001)Harris, Calzetti, III, Conselice, \&
  Smith}]{Harris_2001}
Harris, J., Calzetti, D., III, J. S.~G., Conselice, C.~J., \& Smith, D.~A.
  2001, The Astronomical Journal, 122, 3046

\bibitem[{He {et~al.}(2016)He, Zhang, Ren, \& Sun}]{2015arXiv151203385H}
He, K., Zhang, X., Ren, S., \& Sun, J. 2016, in 2016 IEEE Conference on
  Computer Vision and Pattern Recognition (CVPR), 770

\bibitem[{{Hernandez} {et~al.}(2019){Hernandez}, {Larsen}, {Aloisi}, {Berg},
  {Blair}, {Fox}, {Heckman}, {James}, {Long}, {Skillman}, \&
  {Whitmore}}]{2019ApJ...872..116H}
{Hernandez}, S., {Larsen}, S., {Aloisi}, A., {et~al.} 2019, \apj, 872, 116

\bibitem[{{Kingma} \& {Ba}(2014)}]{2014arXiv1412.6980K}
{Kingma}, D.~P. \& {Ba}, J. 2014, ArXiv e-prints [\eprint[arXiv]{1412.6980}]

\bibitem[{{Krist} {et~al.}(2011){Krist}, {Hook}, \&
  {Stoehr}}]{2011SPIE.8127E..0JK}
{Krist}, J.~E., {Hook}, R.~N., \& {Stoehr}, F. 2011, in \procspie, Vol. 8127,
  Optical Modeling and Performance Predictions V, 81270J

\bibitem[{{Kroupa}(2001)}]{2001MNRAS.322..231K}
{Kroupa}, P. 2001, \mnras, 322, 231

\bibitem[{{Krumholz} {et~al.}(2015){Krumholz}, {Fumagalli}, {da Silva},
  {Rendahl}, \& {Parra}}]{2015MNRAS.452.1447K}
{Krumholz}, M.~R., {Fumagalli}, M., {da Silva}, R.~L., {Rendahl}, T., \&
  {Parra}, J. 2015, \mnras, 452, 1447

\bibitem[{{McConnachie} {et~al.}(2005){McConnachie}, {Irwin}, {Ferguson},
  {Ibata}, {Lewis}, \& {Tanvir}}]{2005MNRAS.356..979M}
{McConnachie}, A.~W., {Irwin}, M.~J., {Ferguson}, A.~M.~N., {et~al.} 2005,
  \mnras, 356, 979

\bibitem[{{Portegies Zwart} {et~al.}(2010){Portegies Zwart}, {McMillan}, \&
  {Gieles}}]{2010ARA&A..48..431P}
{Portegies Zwart}, S.~F., {McMillan}, S.~L.~W., \& {Gieles}, M. 2010, \araa,
  48, 431

\bibitem[{{Reyes} {et~al.}(2018){Reyes}, {Est{\'e}vez}, {Reyes},
  {Cabrera-Vives}, {Huijse}, {Carrasco-Davis}, \&
  {F{\"o}rster}}]{Reyes2018EnhancedRI}
{Reyes}, E., {Est{\'e}vez}, P.~A., {Reyes}, I., {et~al.} 2018, 2018
  International Joint Conference on Neural Networks (IJCNN), 1

\bibitem[{{Ribli} {et~al.}(2019){Ribli}, {Dobos}, \&
  {Csabai}}]{2019arXiv190208161R}
{Ribli}, D., {Dobos}, L., \& {Csabai}, I. 2019, arXiv e-prints,
  arXiv:1902.08161

\bibitem[{{Rowe} {et~al.}(2015){Rowe}, {Jarvis}, {Mandelbaum}, {Bernstein},
  {Bosch}, {Simet}, {Meyers}, {Kacprzak}, {Nakajima}, {Zuntz}, {Miyatake},
  {Dietrich}, {Armstrong}, {Melchior}, \& {Gill}}]{2015A&C....10..121R}
{Rowe}, B.~T.~P., {Jarvis}, M., {Mandelbaum}, R., {et~al.} 2015, Astronomy and
  Computing, 10, 121

\bibitem[{Russakovsky {et~al.}(2015)Russakovsky, Deng, Su, Krause, Satheesh,
  Ma, Huang, Karpathy, Khosla, Bernstein, Berg, \& Fei-Fei}]{ILSVRC15}
Russakovsky, O., Deng, J., Su, H., {et~al.} 2015, International Journal of
  Computer Vision, 115, 211

\bibitem[{{Ryon} {et~al.}(2015){Ryon}, {Bastian}, {Adamo}, {Konstantopoulos},
  {Gallagher}, {Larsen}, {Hollyhead}, {Silva-Villa}, \&
  {Smith}}]{2015MNRAS.452..525R}
{Ryon}, J.~E., {Bastian}, N., {Adamo}, A., {et~al.} 2015, \mnras, 452, 525

\bibitem[{{S{\'a}nchez-Gil} {et~al.}(2019){S{\'a}nchez-Gil}, {Alfaro},
  {Cervi{\~n}o}, {P{\'e}rez}, {Bland -Hawthorn}, \&
  {Jones}}]{2019MNRAS.483.2641S}
{S{\'a}nchez-Gil}, M.~C., {Alfaro}, E.~J., {Cervi{\~n}o}, M., {et~al.} 2019,
  \mnras, 483, 2641

\bibitem[{{Thim} {et~al.}(2003){Thim}, {Tammann}, {Saha}, {Dolphin}, {Sandage},
  {Tolstoy}, \& {Labhardt}}]{2003ApJ...590..256T}
{Thim}, F., {Tammann}, G.~A., {Saha}, A., {et~al.} 2003, \apj, 590, 256

\bibitem[{{Vansevi{\v c}ius} {et~al.}(2009){Vansevi{\v c}ius}, {Kodaira},
  {Narbutis}, {Stonkut{\.e}}, {Brid{\v z}ius}, {Deveikis}, \&
  {Semionov}}]{2009ApJ...703.1872V}
{Vansevi{\v c}ius}, V., {Kodaira}, K., {Narbutis}, D., {et~al.} 2009, \apj,
  703, 1872

\bibitem[{{Whitmore} {et~al.}(2011){Whitmore}, {Chandar}, {Kim}, {Kaleida},
  {Mutchler}, {Stankiewicz}, {Calzetti}, {Saha}, {O'Connell}, {Balick}, {Bond},
  {Carollo}, {Disney}, {Dopita}, {Frogel}, {Hall}, {Holtzman}, {Kimble},
  {McCarthy}, {Paresce}, {Silk}, {Trauger}, {Walker}, {Windhorst}, \&
  {Young}}]{2011ApJ...729...78W}
{Whitmore}, B.~C., {Chandar}, R., {Kim}, H., {et~al.} 2011, \apj, 729, 78

\bibitem[{{Wu} {et~al.}(2019){Wu}, {Wong}, {Rudnick}, {Shabala}, {Alger},
  {Banfield}, {Ong}, {White}, {Garon}, \& {Norris}}]{2019MNRAS.482.1211W}
{Wu}, C., {Wong}, O.~I., {Rudnick}, L., {et~al.} 2019, \mnras, 482, 1211

\end{thebibliography}
\bibliographystyle{aa}
\end{document}